\documentclass[showpacs,aps,graphicx,twocolumn]{revtex4}
\usepackage{amsmath}
\usepackage{graphicx}
\usepackage{verbatim}
\usepackage[caption=false,font=scriptsize,labelfont=sf,textfont=sf]{subfig}
\usepackage{bm}
\allowdisplaybreaks[4]
\begin{document}
\title{Direct generation of multi-photon hyperentanglement}

\author{Peng Zhao$^{1,2}$, Jia-Wei Ying$^{1,2}$, Meng-Ying Yang$^{1,2}$, Wei Zhong$^{2}$, Ming-Ming Du$^{1}$, Shu-Ting Shen$^{1}$, Yun-Xi Li$^{3}$, An-Lei Zhang$^{3}$, Lan Zhou$^{3}$\footnote{Email address: zhoul@njupt.edu.cn}, and Yu-Bo Sheng$^{1,2}$\footnote{Email address: shengyb@njupt.edu.cn}}

\address{$^1$College of Electronic and Optical Engineering \& College of Flexible Electronics (Future
Technology), Nanjing University of Posts and Telecommunications, Nanjing 210023, China\\
 $^2$Institute of Quantum Information and Technology, Nanjing University of Posts and Telecommunications,
Nanjing, 210003, China\\
$^3$College of Science, Nanjing University of Posts and Telecommunications, Nanjing,
210023, China\\ }

\begin{abstract}
  Multi-photon hyperentangement is of fundamental importance in optical quantum information processing. Existing theory and experiment producing multi-photon hyperentangled states have until now relied on the outcome post-selection, a procedure where only the measurement results corresponding to the desired state are considered. Such approach severely limits the usefulness of the resulting hyperentangled states. We present the protocols of direct production of three- and four-photon hyperentanglement and extend the approach to an arbitrary number of photons through a straightforward cascade of spontaneous parametric down-conversion (SPDC) sources. The generated multi-photon hyperentangled states are encoded in polarization-spatial modes and polarization-time bin degrees of freedom, respectively. Numerical calculation shows that if the average photon number $\mu$ is set to 1, the down conversion efficiency is $7.6*10^{-6}$ and the repetition frequency of the laser is $10^9$ Hz, the number of the generation of three-photon and four-photon hyperentanglement after cascading can reach about $5.78*10^{-2}$ and $4.44*10^{-7}$ pairs per second, respectively. By eliminating the constraints of outcome post-selection, our protocols may represent important progresses for multi-photon hyperentangement generation and providing a pivotal role in future multi-party and high-capacity communication networks.
\end{abstract}
\maketitle
\section{Introduction}
Quantum entanglement serves as a fundamental resource for quantum information processing, embodying the core principles of quantum theory, namely coherence and spatial non-locality. These notable characteristics have found widespread applications in quantum communication, including quantum key distribution (QKD) \cite{QKD1}, quantum secret sharing (QSS) \cite{QSS1}, and quantum secure direct communication (QSDC) \cite{QSDC1,QSDC2,onestep}. Numerous methodologies for generating entanglement have been proposed by researchers, and their experimental validations have been undertaken, addressing diverse quantum tasks \cite{entanglement1,entanglement2,entanglement3,entanglement4,entanglement5,entanglement6,entanglement7}.

Researchers have extensively leveraged photons for their rapid transmission speed in quantum entanglement. By exploiting the polarization degree of freedom (DOF) of photons, various communication tasks have been undertaken using polarization Bell states \cite{task2, task3, task4}. Photons possess not only the polarization DOF but also other DOFs, including time-bin, frequency, spatial mode, orbital angular momentum (OAM) and so on. The entanglement of photons in two or more distinct DOFs is termed as hyperentanglement \cite{hyperentangled1,hyperentangled2,hyperentangled3}. Different forms of hyperentanglement, such as polarization-spatial-mode hyperentanglement \cite{PS1,PS2,PS3,PS4}, polarization-frequency hyperentanglement \cite{PF1,PF2,PF3}, and polarization-OAM hyperentanglement \cite{PO1,PO2,PO3,PO4,PO5,PO6}, polarization-time-bin hyperentanglement \cite{PT1,PT2,PT3} have been proposed theoretically and experimentally demonstrated.  Hyperentanglement finds wide applications due to its capacity to enhance channel capacity \cite{High1,High2,onestep}, facilitate comprehensive Bell state analysis \cite{BSM1,BSM2,BSM3,BSM4}, and enable efficient entanglement purification and concentration \cite{purification1,purification2,purification3,purification4,purification5}. Recently, the preparation of  two-photon hyperentanglement in three DOFs were proposed and demonstrated \cite{three1,three2}.

Multi-particle entanglement, such as Greenberger-Horne-Zeilinger (GHZ) state \cite{GHZ1} has found applications in QSS \cite{QSS1}, QSDC \cite{application-qsdc}, quantum teleportation (QT) \cite{application-qt}, and also distributed quantum computation \cite{distributed1}. The generation of multi-particle GHZ states has been realized in ions \cite{ion1,ion2}, photons \cite{photon1,photon2}, and nitrogen-vacancy centers in diamond \cite{NV}, and so on.  Multi-photon hyper-entangled states will also play an important role in increasing the capacity of multi-party quantum communication channels \cite{multihyperentanglement2}, assisting in distinguishing  GHZ states \cite{multihyperentanglement4}, and achieving single-copy entanglement purification \cite{multihyperentanglement5,multihyperentanglement6}. The preparation of six-photon hyperentangled states using nonlinear Kerr media has been proposed \cite{6photon}. Preparing 18-qubit entanglement with six photons encoded in paths, polarization, and orbital angular momentum three DOFs was first reported \cite{multihyperentanglement1}.

Currently, the existing established method for generating photonic hyperentanglement is spontaneous parametric downconversion (SPDC). On the other hand, the experiment with multi-photon hyperentanglement relied on combining photons from two or more different SPDC sources using linear optics and employing outcome post-selection, i. e., the approach that selecting only a specific subset of measurement results while discard others \cite{cascade1,cascade2}. The post-selection approach is the action of observing the photons both creates and destroy the entangled states at the same time, which will restrict the further usefulness. Creating multiphoton hyperentanglement without post-selection would provide a significant advance in photonic quantum communication and quantum network.

In this paper, we propose two protocols to generate multi-photon hyperentangled GHZ states using cascade approach instead of post-selection \cite{cascade1,cascade2}. The first multi-photon hyperentangled GHZ states is encoded in  polarization-spatial-mode DOFs and the second one is encoded in polarization-time-bin DOFs. The structure of this paper is as follows: In Sec. II, we demonstrate the preparation of three-photon polarization-spatial-mode hyperentangled GHZ states. In Sec. III, we extend the preparation method to four-photon polarization-spatial-mode hyperentangled GHZ states and further generalizes it to $m$-photon case. In Sec. IV, we showcase the preparation of three-photon polarization-time-bin hyper-entangled GHZ states. In Sec. V, we extend the methodology to four-photon and $m$-photon polarization-time-bin multi-photon hyperentangled GHZ states. Finally, in Sec. VI, we provide a discussion and conclusion. We also give a detailed calculation of the number of hyperentanglement generated by multi-photon events in the appendix.

\section{Generation of three-photon hyperentanglement in polarization-spatial-mode DOFs}
In this section, we will provide a detailed introduction to the generation method of three-photon hyperentanglement in the polarization-spatial-mode. Before delving into the specifics, let's briefly elucidate the principle of Sagnac interferometer-based polarization entanglement generation \cite{Sagnac}. Upon traversing a polarization beam splitter (PBS), the pump light bifurcates into both directions of the periodically poled potassium titanyl phosphate (ppKTP) crystal within the Sagnac interferometer. The horizontally polarized component $|H\rangle$ pumps in the counterclockwise (CCW) direction, while the vertically polarized component $|V\rangle$ pumps in the clockwise (CW) direction. Subsequently, they reunite in the PBS to engender entangled states in polarization.

 \begin{figure}[!h]
   \centering
  \includegraphics[scale=0.4,angle=0]{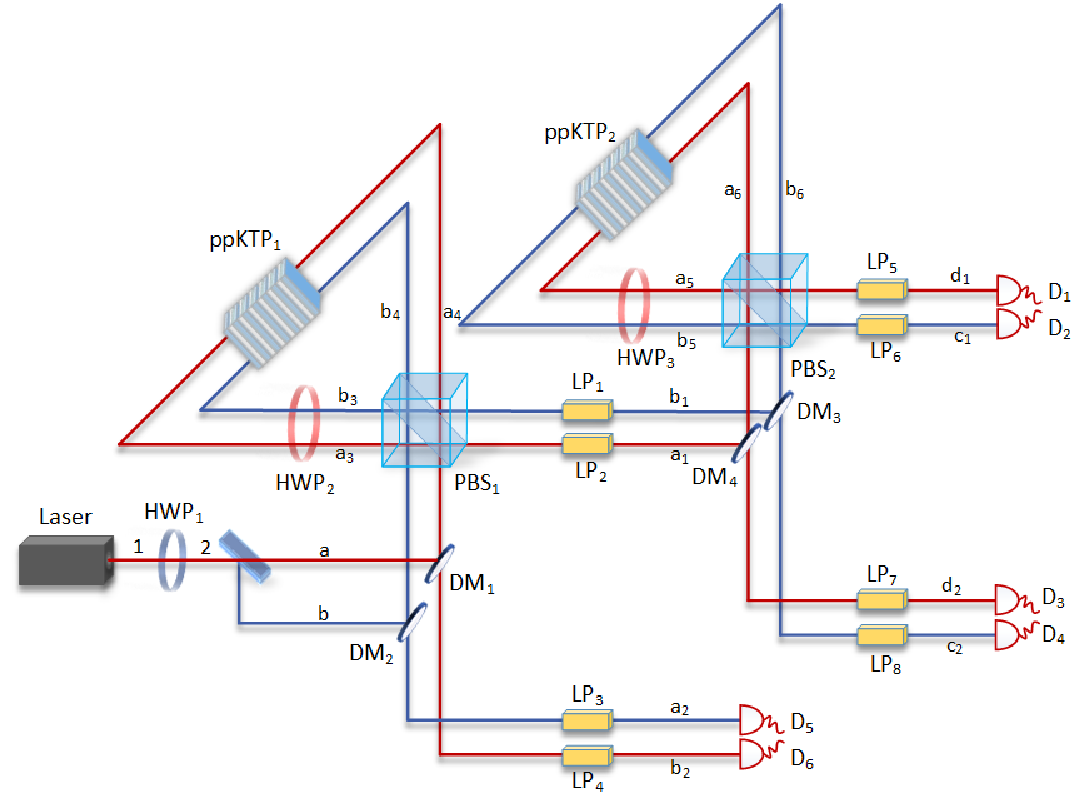}
  \caption{Schematic diagram illustrating the generation of three-photon hyperentanglement in polarization-spatial-mode DOFs. HWP$_1$: $22.5^{\circ}$ half wave plate; HWP$_{2}$ and HWP$_{3}$: $45^{\circ}$ half wave plate; NPBS: non-polarizing beam splitter; DM: dichroic mirror; PBS: polarizing beam splitter; ppKTP: periodically poled potassium titanyI phosphate; LP: long pass filter.}
\end{figure}
(1) As shown in Fig. 1, the laser generates a stream of horizontally polarized pump light $|H\rangle_1$, subject to transformation into diagonally polarized light $\frac{1}{\sqrt{2}}(|H\rangle+|V\rangle)_2$ via a preset angle of $22.5^{\circ}$ with the aid of half wave plate (HWP$_1$). Subsequently, a 50:50 non-polarizing beam splitter (NPBS) divides the photons along distinct trajectories. That is
\begin{eqnarray}\label{1}
  &&\frac{1}{\sqrt{2}}(|H\rangle+|V\rangle)_2 \nonumber\\
  \xrightarrow{NPBS} &&\frac{1}{2}(|H\rangle_a+|H\rangle_b+|V\rangle_a+|V\rangle_b).
\end{eqnarray}

(2) Photons from path a or b enter the Sagnac interferometer through dichroic mirrors (DM$_1$) and DM$_2$ followed by PBS$_1$. They propagate either CW or CCW, subsequently traversing the ppKTP crystal and HWP$_2$ set at a fixed angle of $45^{\circ}$ to generate a two-photon state. HWP$_2$ facilitates the conversion between horizontal polarization and vertical polarization ($|H\rangle \rightleftharpoons |V\rangle$). Consequently, the CW photon passes through ppKTP and then HWP$_2$, while the CCW photon traverses HWP$_2$ and then ppKTP. We can represent this process as follows.
\begin{eqnarray}
  |H\rangle_a &&\xrightarrow{PBS_1} |H\rangle_{a_{4}} \xrightarrow{ppKTP_1+HWP_2} |V\rangle_{a_{3}}|H\rangle_{a_{3}} \nonumber\\
  &&\xrightarrow{PBS_1} |V\rangle_{a_2}|H\rangle_{a_1},  \nonumber\\
  |V\rangle_a &&\xrightarrow{PBS_1} |V\rangle_{a_{3}} \xrightarrow{HWP_2+ppKTP_1} |H\rangle_{a_{4}}|V\rangle_{a_{4}} \nonumber\\
  &&\xrightarrow{PBS_1} |H\rangle_{a_2}|V\rangle_{a_1},  \nonumber\\
  |H\rangle_b &&\xrightarrow{PBS_1} |H\rangle_{b_{4}} \xrightarrow{ppKTP_1+HWP_2} |V\rangle_{b_{3}}|H\rangle_{b_{3}} \nonumber\\
  &&\xrightarrow{PBS_1} |V\rangle_{b_2}|H\rangle_{b_1},  \nonumber\\
  |V\rangle_b &&\xrightarrow{PBS_1} |V\rangle_{b_{3}} \xrightarrow{HWP_2+ppKTP_1} |H\rangle_{b_{4}}|V\rangle_{b_{4}} \nonumber\\
  &&\xrightarrow{PBS_1} |H\rangle_{b_2}|V\rangle_{b_1}.
\end{eqnarray}

(3) The photons, arriving at PBS$_1$ for the second time in both CW and CCW directions, undergo path differentiation through PBS$_1$. Consequently, Eq. (\ref{1}) ultimately evolves into Eq. (\ref{3}) after traversing PBS$_1$.
\begin{eqnarray}\label{3}
&&\frac{1}{2}(|H\rangle_a+|H\rangle_b+|V\rangle_a+|V\rangle_b) \nonumber\\
   &\xrightarrow{PBS_1}& \frac{1}{2}(|V\rangle_{a_2}|H\rangle_{a_1}+|H\rangle_{a_2}|V\rangle_{a_1}+|V\rangle_{b_2}|H\rangle_{b_1} \nonumber\\
   &+& |H\rangle_{b_2}|V\rangle_{b_1}) \nonumber\\
   &=& \frac{1}{2}(|HV\rangle+|VH\rangle)\otimes(|a_1a_2\rangle+|b_1b_2\rangle).
\end{eqnarray}

(4) In Eq. (\ref{3}), it is evident that we have achieved polarization-spatial-mode two-photon hyperentanglement. Subsequently, by directing one of the particles into a Sagnac interferometer, successful photon splitting will yield polarization-spatial hyperentanglement among three photons. Specifically, we connect the Sagnac interferometer to paths $a_1$ and $b_1$. The states described in Eq. (\ref{3}) will initially pass through PBS$_2$ and undergo transformation into the new state as
\begin{eqnarray}\label{4}
&&\frac{1}{2}(|HV\rangle+|VH\rangle)\otimes(|a_1a_2\rangle+|b_1b_2\rangle) \nonumber\\
&=&\frac{1}{2}(|V\rangle_{a_2}|H\rangle_{a_1}+|H\rangle_{a_2}|V\rangle_{a_1}+|V\rangle_{b_2}|H\rangle_{b_1} \nonumber\\
&&+|H\rangle_{b_2}|V\rangle_{b_1}) \nonumber\\
&\xrightarrow{PBS_2}&\frac{1}{2}(|V\rangle_{a_2}|H\rangle_{a_6}+|H\rangle_{a_2}|V\rangle_{a_5}+|V\rangle_{b_2}|H\rangle_{b_6} \nonumber\\
&&+|H\rangle_{b_2}|V\rangle_{b_5}).
\end{eqnarray}

(5) Analogously, the quantum states traveling in both CW and CCW directions will experience the subsequent transformation within the Sagnac interferometer. That is
\begin{eqnarray}\label{5}
  &&|H\rangle_{a_6} \xrightarrow{ppKTP_2+HWP_3} |V\rangle_{a_{5}}|H\rangle_{a_{5}}\xrightarrow{PBS_1} |V\rangle_{d_2}|H\rangle_{d_1}, \nonumber\\
  &&|V\rangle_{a_5} \xrightarrow{HWP_3+ppKTP_2} |H\rangle_{a_{6}}|V\rangle_{a_{6}}\xrightarrow{PBS_1} |H\rangle_{d_2}|V\rangle_{d_1},  \nonumber\\
  &&|H\rangle_{b_6} \xrightarrow{ppKTP_2+HWP_3} |V\rangle_{b_{5}}|H\rangle_{b_{5}}\xrightarrow{PBS_1} |V\rangle_{c_2}|H\rangle_{c_1},  \nonumber\\
  &&|V\rangle_{b_5} \xrightarrow{HWP_3+ppKTP_2} |H\rangle_{b_{6}}|V\rangle_{b_{6}}\xrightarrow{PBS_1} |H\rangle_{c_2}|V\rangle_{c_1}. \nonumber\\
\end{eqnarray}

(6) The photons, traversing in both CW and CCW directions, re-enter PBS$_2$ for the second time and undergo path separation. By combining the expressions in Eqs. (\ref{4}) and (\ref{5}), we can obtain
\begin{eqnarray}
&&\frac{1}{2}(|V\rangle_{a_2}|H\rangle_{a_6}+|H\rangle_{a_2}|V\rangle_{a_5}+|V\rangle_{b_2}|H\rangle_{b_6} \nonumber\\
&&+|H\rangle_{b_2}|V\rangle_{b_5}) \nonumber\\
&\xrightarrow{PBS_2}&\frac{1}{2}(|V\rangle_{a_2}|V\rangle_{d_2}|H\rangle_{d_1}+|H\rangle_{a_2}|H\rangle_{d_2}|V\rangle_{d_1} \nonumber\\
&& +|V\rangle_{b_2}|V\rangle_{c_2}|H\rangle_{c_1}+|H\rangle_{b_2}|H\rangle_{c_2}|V\rangle_{c_1}) \nonumber\\
&=&\frac{1}{2}(|HVH\rangle+|VHV\rangle)\otimes(|c_2c_1b_2\rangle+|d_2d_1a_2\rangle). \nonumber\\
\end{eqnarray}

(7) Ultimately, we employ six long pass filters (LPs) in modes a$_2$, b$_2$, c$_1$, d$_1$, c$_2$, d$_2$ to filter the three-photon hyperentanglement in polarization-spatial-mode DOFs as shown in above equation.

\section{Generation of four-photon and $m$-photon hyperentanglement in polarization-spatial-mode DOFs}
\begin{figure}[!htp]
  \centering
 \includegraphics[scale=0.4,angle=0]{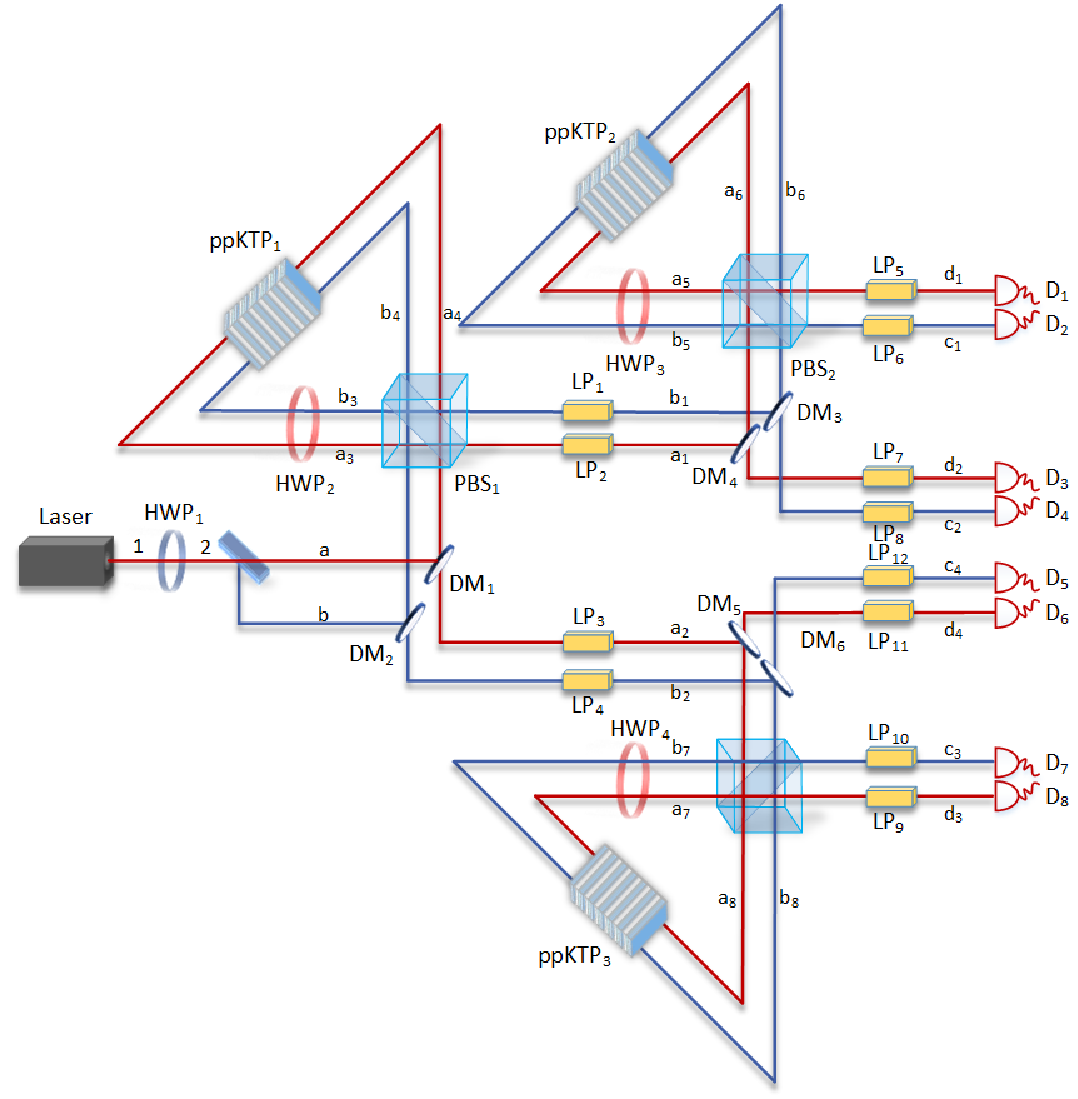}
  \caption{Schematic diagram illustrating the generation of four-photon hyperentanglement in polarization-spatial-mode DOFs. HWP$_1$: $22.5^{\circ}$ half wave plate; HWP$_{2}$, HWP$_{3}$ and HWP$_{4}$: $45^{\circ}$ half wave plate; NPBS: non-polarizing beam splitter; DM: dichroic mirror; PBS: polarizing beam splitter; ppKTP: periodically poled potassium titanyI phosphate; LP: long pass filter.}
\end{figure}

The symmetry of the structure becomes apparent when generating four-photon hyperentanglement. As shown in Fig. 2, the utilization of photons in paths a$_2$ and b$_2$ allows for the creation of four-photon hyperentanglement in polarization-spatial-mode DOFs. It is noteworthy that the initial three steps in generating four-photon hyperentanglement mirror those in generating three-photon hyperentanglement, establishing a seamless transition. For continuity, let's commence with Eq. (\ref{3}).

(1) Firstly, we suppose that we have generated two-photon hyperentanglement in the spatial modes a$_1$ and a$_2$, b$_1$ and b$_2$. When photons from spatial modes a$_1$ and a$_2$, b$_1$ and b$_2$ initially enter PBS$_2$ and PBS$_3$, respectively, Eq. (\ref{3}) can be evolved as
\begin{eqnarray}\label{7}
&&\frac{1}{2}(|HV\rangle+|VH\rangle)\otimes(|a_1a_2\rangle+|b_1b_2\rangle) \nonumber\\
&\xrightarrow{PBS_2+PBS_3}&\frac{1}{2}(|V\rangle_{a_7}|H\rangle_{a_6}+|H\rangle_{a_8}|V\rangle_{a_5}+|V\rangle_{b_7}|H\rangle_{b_6} \nonumber\\
&&+|H\rangle_{b_8}|V\rangle_{b_5}).
\end{eqnarray}

\begin{figure}[!htp]
  \centering
 \includegraphics[scale=0.5,angle=0]{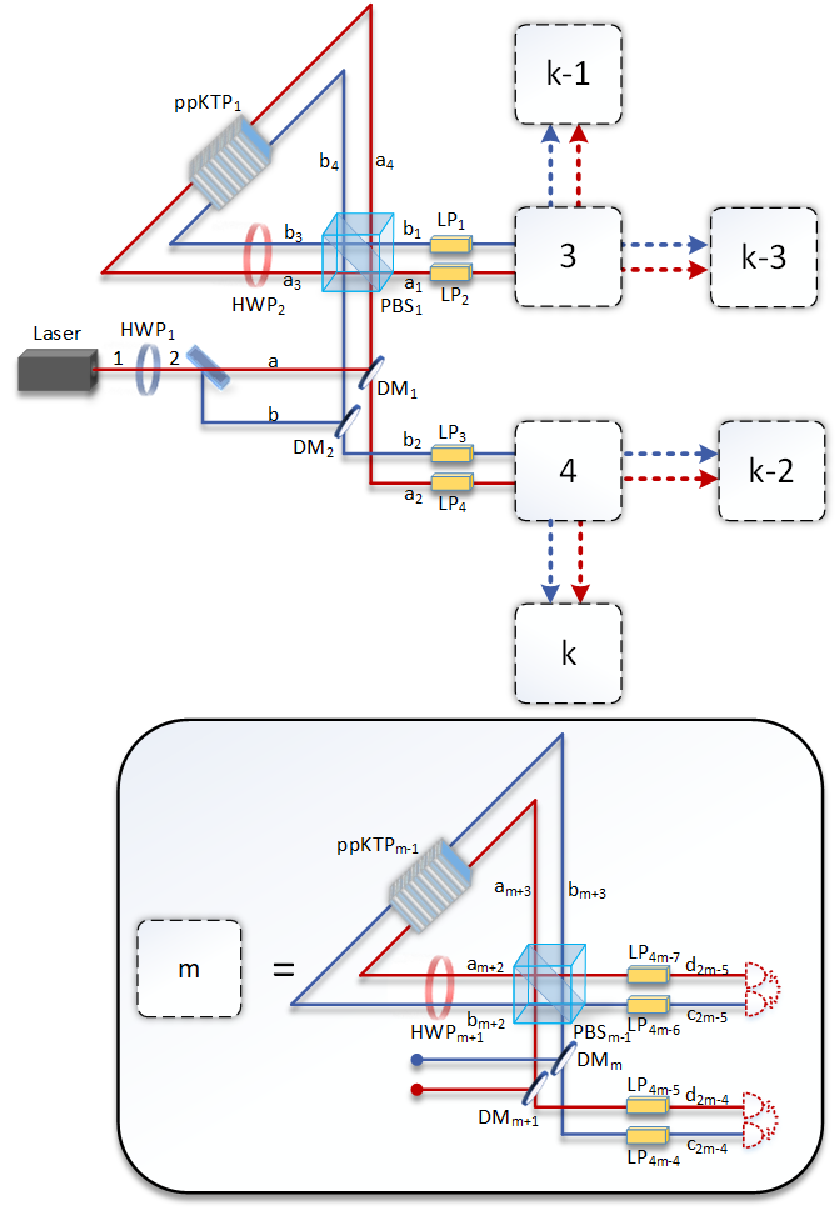}
  \caption{Schematic diagram illustrating the generation of $m$-photon hyperentanglement in polarization-spatial-mode DOFs. HWP$_1$: $22.5^{\circ}$ half wave plate; HWP$_{2}$: $45^{\circ}$ half wave plate; NPBS: non-polarizing beam splitter; DM: dichroic mirror; PBS: polarizing beam splitter; ppKTP: periodically poled potassium titanyI phosphate; LP: long pass filter. In the legend, $m = 3, 4, \cdots k$ is the scenario for preparing $m$-photon hyperentanglement in polarization-spatial-mode DOFs. a$_{m+2}$, b$_{m+2}$, a$_{m+3}$, b$_{m+3}$, c$_{m-1}$, d$_{m-1}$, c$_{m-2}$, d$_{m-2}$ represent the corresponding spatial modes after cascading. ppKTP$_{m-1}$ indicates the necessity for $m-1$ ppKTP crystals in the $m$-photon scenario. The red dashed detector indicates that only the final cascade generates photons that may require detection.}
\end{figure}

(2) Analogously, the quantum states in the CW and CCW directions will experience the subsequent transformation within the introduced Sagnac interferometer. That is

\begin{eqnarray}\label{8}
  &&|H\rangle_{a_8}  \xrightarrow{ppKTP_2+HWP_3} |V\rangle_{a_{7}}|H\rangle_{a_{7}} \xrightarrow{PBS_3} |V\rangle_{d_{4}}|H\rangle_{d_{3}},  \nonumber\\
  &&|V\rangle_{a_7}  \xrightarrow{HWP_3+ppKTP_2} |H\rangle_{a_{8}}|V\rangle_{a_{8}} \xrightarrow{PBS_3} |H\rangle_{d_{4}}|V\rangle_{d_{3}},  \nonumber\\
  &&|H\rangle_{b_8}  \xrightarrow{ppKTP_2+HWP_1} |V\rangle_{b_{7}}|H\rangle_{b_{7}} \xrightarrow{PBS_3} |V\rangle_{c_{4}}|H\rangle_{c_{3}},  \nonumber\\
  &&|V\rangle_{b_7}  \xrightarrow{HWP_3+ppKTP_2} |H\rangle_{b_{8}}|V\rangle_{b_{8}} \xrightarrow{PBS_3} |H\rangle_{c_{4}}|V\rangle_{c_{3}}. \nonumber\\
\end{eqnarray}

(3) After propagating CW or CCW in the Sagnac interferometer, the photons in paths a$_5$, a$_6$, b$_5$, and b$_6$ are separated into different paths upon re-entering PBS$_2$ for the second time. The same principle applies to the photons on paths a$_7$, a$_8$, b$_7$, and b$_8$. Thus, in conjunction with Eqs. (\ref{7}) and (\ref{8}), we can obtain the state in Eq. (\ref{4hyper}). This represents the  four-photon hyperentanglement in polarization-spatial-mode DOFs as

\begin{eqnarray}\label{4hyper}
&&\frac{1}{2}(|V\rangle_{a_7}|H\rangle_{a_6}+|H\rangle_{a_8}|V\rangle_{a_5}+|V\rangle_{b_7}|H\rangle_{b_6} \nonumber\\
&&+|H\rangle_{b_8}|V\rangle_{b_5}) \nonumber\\
&\xrightarrow{PBS_2+PBS_3}&\frac{1}{2}(|H\rangle_{d_4}|V\rangle_{d_3}|V\rangle_{d_2}|H\rangle_{d_1}+|V\rangle_{d_4}|H\rangle_{d_3} \nonumber\\
&&\otimes |H\rangle_{d_2}|V\rangle_{d_1} +|H\rangle_{c_4}|V\rangle_{c_3}|V\rangle_{c_2}|H\rangle_{c_1} \nonumber\\
&&+|V\rangle_{c_4}|H\rangle_{c_3}|V\rangle_{c_2}|V\rangle_{c_1}) \nonumber\\
&=&\frac{1}{2}(|HVVH\rangle+|VHHV\rangle)\otimes(|d_4d_3d_2d_1\rangle \nonumber\\
&&+|c_4c_3c_2c_1\rangle).
\end{eqnarray}

(4) Finally, we employ LPs to eliminate residual pump light and background photons, ensuring the acquisition of  four-photon hyperentanglement in polarization-spatial-mode DOFs.

So far, we have provided comprehensive details on the preparation of three- and four-photon hyperentanglements in polarization-spatial-mode DOFs. Subsequently, we will briefly describe the scenario for preparing $m$-photon hyperentanglement. From Fig. 3, it is evident that the theoretical achievement of $m$-photon hyperentanglement is feasible through this straightforward cascaded SPDC scheme. The legend in Fig. 3 explains the meaning of the dashed box, where $m$ = $ 3, 4, \cdots k$, denotes the scenario of preparing $m$-photon hyperentanglement in polarization-spatial-mode DOFs. a$_{m+2}$, b$_{m+2}$, a$_{m+3}$, b$_{m+3}$, c$_{m-1}$, d$_{m-1}$, c$_{m-2}$, d$_{m-2}$ mean the corresponding spatial modes after cascading. ppKTP$_{m-1}$ indicates the requirement for $m-1$ ppKTP crystals in the $m$-photon scenario. Without loss of generality, we assume that $m$ is even. The desired $m$-photon hyperentanglement in polarization-spatial-mode DOFs as given in Eq. (\ref{desired}) can be generated.

\begin{eqnarray}\label{desired}
 &&\frac{1}{(\sqrt{2})^{m}}(|HV\ldots VH\rangle+|VH\ldots HV\rangle) \nonumber\\
 &&\otimes(|d_{2m-4}d_{2m-3}\ldots d_2d_1\rangle+|c_{2m-4}c_{2m-3}\ldots c_2c_1\rangle). \nonumber\\
\end{eqnarray}

It is noteworthy that in practical laser sources, multi-photon events are inevitable. Investigating the impact of multi-photon events on hyperentanglement generation is crucial. Fortunately, researchers have already explored the influence of multi-photon events on the generation of hyperentanglement in three DOFs, and we will not delve into it in this paper. For the proposed scheme of generating multi-photon hyperentanglement in this study, we will provide a detailed discussion of the efficiency of multi-photon events in the appendix.

\section{Generation of three-photon hyperentanglement in polarization-time-bin DOFs}
In this section, we describe the protocol to generate the three-photon, four-photon and $m$-photon hyperentanglements in polarization-time-bin DOFs, respectively. We first describe the generation of three-photon hyperentanglement as follows:

(1) As shown in Fig. 4, the laser generates a beam of horizontally polarized pump light $|H\rangle_1$, which is converted into diagonally polarized light $\frac{1}{\sqrt{2}}(|Ht_1\rangle+|Ht_2\rangle)$ through a 50:50 NPBS$_1$. Here $t_{1}$ and $t_{2}$ mean that the photon is in the short and long arm, respectively.  Subsequently, the state undergoes a transformation as follows.
\begin{eqnarray}
   &&\frac{1}{\sqrt{2}}(|Ht_1\rangle+|Ht_2\rangle) \nonumber\\
\xrightarrow{NPBS_2}&&\frac{1}{2}(|Ht_1\rangle_2+|Ht_1\rangle_c+|Ht_2\rangle_2-|Ht_2\rangle_c).
\end{eqnarray}
Here we focus only on path 2, omitting path c for simplicity. The photon in path 2 passes through a $22.5^{\circ}$ HWP$_1$, resulting in
\begin{eqnarray}\label{HWP1}
 &&\frac{1}{\sqrt{2}}(|Ht_1\rangle_2+|Ht_2\rangle_2) \nonumber\\
  \xrightarrow{HWP_1}&&\frac{1}{2}(|Ht_1\rangle_3+|Vt_1\rangle_3+|Ht_2\rangle_3+|Vt_2\rangle_3).
\end{eqnarray}

\begin{figure}[!htp]
  \centering
 \includegraphics[scale=0.3,angle=0]{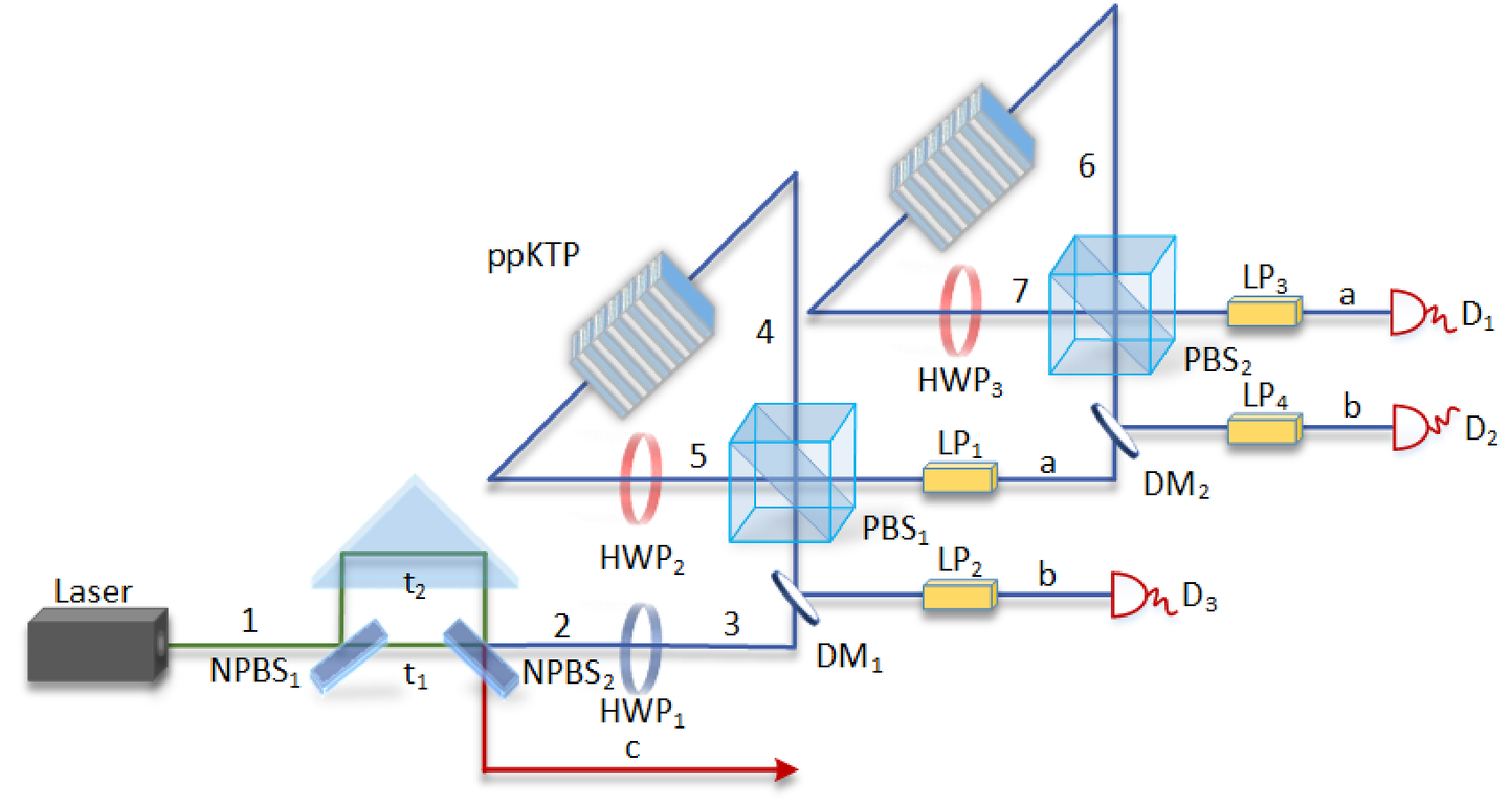}
  \caption{The schematic diagram illustrates the generation of three-photon hyperentanglement in polarization-time-bin DOFs.  HWP$_1$: $22.5^{\circ}$ half wave plate; HWP$_{2}$ and HWP$_{3}$: $45^{\circ}$ half wave plate; NPBS: non-polarizing beam splitter; DM: dichroic mirror; PBS: polarizing beam splitter; ppKTP: periodically poled potassium titanyI phosphate; LP: long pass filter.}
\end{figure}

(2) The polarized photon in spatial mode $3$ successively passes through the DM, PBS, polarization Sagnac interferometer, and PBS to generate the polarization-time-bin hyperentangled photon pair in the SPDC process. Specifically, the evolution process of the state can be written as
\begin{eqnarray}
  |Ht_1\rangle_3 &&\xrightarrow{PBS_1} |Ht_1\rangle_4 \xrightarrow{ppKTP_1+HWP_2}|Vt_1\rangle_{5}|Ht_1\rangle_{5}    \nonumber\\
  &&\xrightarrow{PBS_1} |Vt_1\rangle_{b}|Ht_1\rangle_{a},  \nonumber\\
  |Vt_1\rangle_3 &&\xrightarrow{PBS_1} |Vt_1\rangle_5 \xrightarrow{HWP_2+ppKTP_1} |Ht_1\rangle_{4}|Vt_1\rangle_{4} \nonumber\\
  &&\xrightarrow{PBS_1} |Ht_1\rangle_{b}|Vt_1\rangle_{a},  \nonumber\\
  |Ht_2\rangle_3 &&\xrightarrow{PBS_1} |Ht_2\rangle_4 \xrightarrow{ppKTP_1+HWP_2} |Vt_2\rangle_{5}|Ht_2\rangle_{5} \nonumber\\
  &&\xrightarrow{PBS_1} |Vt_2\rangle_{a}|Ht_2\rangle_{b},  \nonumber\\
  |Vt_2\rangle_3 &&\xrightarrow{PBS_1} |Vt_2\rangle_5 \xrightarrow{HWP_2+ppKTP_1} |Ht_2\rangle_{4}|Vt_2\rangle_{4} \nonumber\\
  &&\xrightarrow{PBS_1} |Ht_2\rangle_{a}|Vt_2\rangle_{b}.
\end{eqnarray}
In this manner, Eq. (\ref{HWP1}) ultimately transforms into Eq. (\ref{12}) after passing through PBS$_1$.
\begin{eqnarray}\label{12}
&&\frac{1}{2}(|Ht_1\rangle_3+|Vt_1\rangle_3+|Ht_2\rangle_3+|Vt_2\rangle_3) \nonumber\\
   &\xrightarrow{PBS_1}& \frac{1}{2}(|Vt_1\rangle_{b}|Ht_1\rangle_{a}+|Ht_1\rangle_{b}|Vt_1\rangle_{a}+|Vt_2\rangle_{a}|Ht_2\rangle_{b} \nonumber\\
   &&+|Ht_2\rangle_{a}|Vt_2\rangle_{b}) \nonumber\\
   &=& \frac{1}{2}(|HV\rangle+|VH\rangle)_{ab}\otimes(|t_1t_1\rangle+|t_2t_2\rangle)_{ab}.
\end{eqnarray}

(3) From Eq. (\ref{12}), it is actually a polarization-time-bin two-photon hyperentanglement \cite{Kwiat}. Subsequently, by directing one of photons into a Sagnac interferometer, successful photon splitting will yield polarization-time-bin hyperentanglement among three photons. Specifically, we cascade  another Sagnac interferometer on paths $a$ and $b$. States in Eq. (\ref{12}) will first pass PBS$_2$ and become
\begin{eqnarray}\label{15}
&&\frac{1}{2}(|HV\rangle+|VH\rangle)_{ab}\otimes(|t_1t_1\rangle+|t_2t_2\rangle)_{ab} \nonumber\\
\xrightarrow{PBS_2}&&\frac{1}{2}(|Vt_1\rangle_{b}|Ht_1\rangle_{6}+|Ht_1\rangle_{b}|Vt_1\rangle_{7}+|Vt_2\rangle_{7}|Ht_2\rangle_{b} \nonumber\\
&&+|Ht_2\rangle_{6}|Vt_2\rangle_{b}).
\end{eqnarray}
Similarly, the CW and CCW quantum states will undergo the following transformation in the Sagnac interferometer.

\begin{eqnarray}\label{16}
 &&|Ht_1\rangle_{6} \xrightarrow{ppKTP_2+HWP_3} |Vt_1\rangle_{7}|Ht_1\rangle_{7} \nonumber\\
 &\xrightarrow{PBS_2}& |Vt_1\rangle_{b_1}|Ht_1\rangle_{a_1},  \nonumber\\
 &&|Vt_1\rangle_{7} \xrightarrow{HWP_3+ppKTP_2} |Ht_1\rangle_{6}|Vt_1\rangle_{6} \nonumber\\
 &\xrightarrow{PBS_2}& |Ht_1\rangle_{b_1}|Vt_1\rangle_{a_1},  \nonumber\\
 &&|Ht_2\rangle_{6} \xrightarrow{ppKTP_2+HWP_3} |Vt_2\rangle_{7}|Ht_2\rangle_{7} \nonumber\\
 &\xrightarrow{PBS_2}& |Vt_2\rangle_{b_1}|Ht_2\rangle_{a_1},  \nonumber\\
 &&|Vt_2\rangle_{7} \xrightarrow{HWP_3+ppKTP_2} |Ht_2\rangle_{6}|Vt_2\rangle_{6} \nonumber\\
 &\xrightarrow{PBS_2}& |Ht_2\rangle_{b_1}|Vt_2\rangle_{a_1}.
\end{eqnarray}

(4) Photons in the CW and CCW directions then enter the PBS$_2$  for the second time. In this way, states in Eqs. (\ref{15}) and (\ref{16}) will evolve into
\begin{eqnarray}\label{17}
&&\frac{1}{2}(|Vt_1\rangle_{b}|Ht_1\rangle_{6}+|Ht_1\rangle_{b}|Vt_1\rangle_{7}+|Vt_2\rangle_{7}|Ht_2\rangle_{b} \nonumber\\
&&+|Ht_2\rangle_{6}|Vt_2\rangle_{b}) \nonumber\\
\rightarrow &&\frac{1}{2}(|Vt_1\rangle_{b}|Vt_1\rangle_{b_1}|Ht_1\rangle_{a_1}+|Ht_1\rangle_{b}|Ht_1\rangle_{b_1}|Vt_1\rangle_{a_1} \nonumber\\
&&+|Ht_2\rangle_{b_1}|Vt_2\rangle_{a_1}|Ht_2\rangle_{b} +|Vt_2\rangle_{b_1}|Ht_2\rangle_{a_1}|Vt_2\rangle_{b}) \nonumber\\
&=&\frac{1}{2}[(|HVH\rangle+|VHV\rangle)\otimes(t_1t_1t_1+t_2t_2t_2)]_{b_1a_1b}, \nonumber\\
\end{eqnarray}
which is the target three-photon hyperentanglement in polarization-time-bin DOFs.

(5) LPs are also utilized to filter out the remaining pump light and background photons, ensuring the attainment of a pure target state.

\section{Generation of four-photon and $m$-photon hyperentanglements in polarization-time-bin DOFs}

\begin{figure}[htp]
  \centering
 \includegraphics[scale=0.3,angle=0]{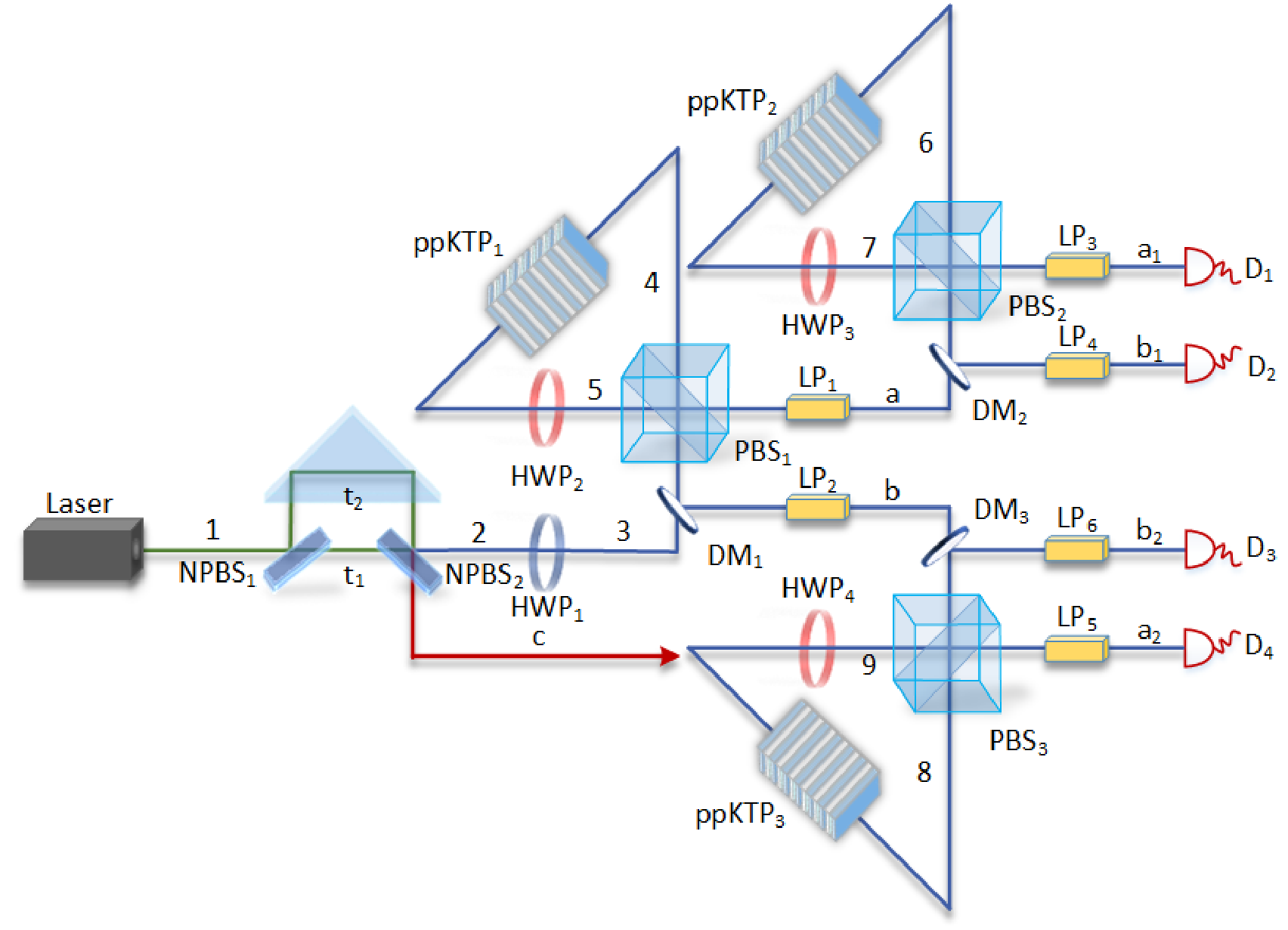}
  \caption{The schematic illustrates the generation of four-photon hyperentanglement in polarization-time-bin DOFs. HWP$_1$: $22.5^{\circ}$ half wave plate; HWP$_{2}$, HWP$_{3}$ and HWP$_{4}$: $45^{\circ}$ half wave plate; NPBS: non-polarizing beam splitter; DM: dichroic mirror; PBS: polarizing beam splitter; ppKTP: periodically poled potassium titanyI phosphate; LP: long pass filter.}
\end{figure}

The generation of four-photon hyperentanglement in polarization-time-bin DOFs relies on simultaneously utilizing the photons generated in Fig. 4. For clarity and convenience, we begin with Eq. (\ref{12}).

(1) Firstly, as shown in Fig. 5, we suppose that we generated the  two-photon hyperentanglement in polarization-time-bin DOFs like ref. \cite{three2}. As photons from paths $a$ and $b$ initially enter PBS$_2$ and PBS$_3$, respectively, states in the Eq. (\ref{12}) undergoes evolution, resulting into
\begin{eqnarray}\label{PBS23}
&&\frac{1}{2}(|HV\rangle+|VH\rangle)_{ab}\otimes(|t_1t_1\rangle+|t_2t_2\rangle)_{ab} \nonumber\\
&\xrightarrow{PBS_2+PBS_3}&\frac{1}{2}(|Ht_1\rangle_{6}|Vt_1\rangle_{9}+|Ht_2\rangle_{6}|Vt_2\rangle_{9} \nonumber\\
&&+|Vt_1\rangle_{7}|Ht_1\rangle_{8}+|Vt_2\rangle_{7}|Ht_2\rangle_{8}).
\end{eqnarray}

(2) The polarized photons in the added Sagnac interferometer will undergo the following transformations when the photons  go through either CW or CCW.
\begin{eqnarray}\label{ppKTP23}
  &&|Ht_1\rangle_{8} \xrightarrow{ppKTP_3+HWP_4} |Vt_1\rangle_{9}|Ht_1\rangle_{9} \nonumber\\
  &\xrightarrow{PBS_3}& |Vt_1\rangle_{b_2}|Ht_1\rangle_{a_2},  \nonumber\\
  &&|Vt_1\rangle_{9} \xrightarrow{HWP_4+ppKTP_3} |Ht_1\rangle_{8}|Vt_1\rangle_{8} \nonumber\\
  &\xrightarrow{PBS_3}& |Ht_1\rangle_{b_2}|Vt_1\rangle_{a_2},  \nonumber\\
  &&|Ht_2\rangle_{8} \xrightarrow{ppKTP_3+HWP_4} |Vt_2\rangle_{9}|Ht_2\rangle_{9} \nonumber\\
  &\xrightarrow{PBS_3}& |Vt_2\rangle_{b_2}|Ht_2\rangle_{a_2},  \nonumber\\
  &&|Vt_2\rangle_{9} \xrightarrow{HWP_4+ppKTP_3} |Ht_2\rangle_{8}|Vt_2\rangle_{8} \nonumber\\
  &\xrightarrow{PBS_3}& |Ht_2\rangle_{b_2}|Vt_2\rangle_{a_2}.
\end{eqnarray}

(3) Photons in spatial modes 6, 7, 8, and 9 will then enter in PBS$_2$ and PBS$_3$ for the second time. Afterwards, photons are separated into different spatial modes by PBS$_2$ and PBS$_3$. In this way, combined with Eqs. (\ref{PBS23}) and (\ref{ppKTP23}) we can obtain
\begin{eqnarray}
&&\frac{1}{2}(|Ht_1\rangle_{6}|Vt_1\rangle_{9}+|Ht_2\rangle_{6}|Vt_2\rangle_{9}+|Vt_1\rangle_{7}|Ht_1\rangle_{8} \nonumber\\
&&+|Vt_2\rangle_{7}|Ht_2\rangle_{8})  \nonumber\\
\rightarrow &&\frac{1}{2}(|Ht_1\rangle_{b_2}|Vt_1\rangle_{a_2}|Vt_1\rangle_{b_1}|Ht_1\rangle_{a_1}+|Vt_1\rangle_{b_2}|Ht_1\rangle_{a_2} \nonumber\\
&&\otimes|Ht_1\rangle_{b_1}|Vt_1\rangle_{a_1}+|Ht_2\rangle_{b_1}|Vt_2\rangle_{a_1}|Vt_2\rangle_{b_2}|Ht_2\rangle_{a_2}  \nonumber\\ &&+|Vt_2\rangle_{b_1}|Ht_2\rangle_{a_1}|Ht_2\rangle_{b_2}|Vt_2\rangle_{a_2}) \nonumber\\
&&=\frac{1}{2}[(|HVVH\rangle +|VHHV\rangle)\otimes(t_1t_1t_1t_1 \nonumber\\
&&+t_2t_2t_2t_2)]_{b_1a_1b_2a_2},
\end{eqnarray}
which is the target four-photon hyperentanglement in polarization-time-bin DOFs.

(4) Finally, it is imperative to utilize LPs to eliminate the pump light and background photons in each mode.

So far, we have provided detailed explanations for three- and four-photon hyperentanglements in polarization-time-bin DOFs. Next, we briefly describe the scenario for preparing $m$-photon hyperentanglement. As depicted in Fig. 6, this can be theoretically achieved by cascading SPDCs. The legend in Fig. 6 explains the meaning of the black dashed box. Here, $m = 3, 4, \cdots k$ represent the devices that need to be added to generate $m$-photon hyperentanglement in polarization-time-bin DOFs. $m+3$, $m+4$, a$_{m-2}$, b$_{m-2}$ represent the path labels. The subscripts $m$ and $m-1$ of HWP$_m$, ppKTP$_{m-1}$, PBS$_{m-1}$, DM$_{m-1}$ represent the number of optical devices required to generate $m$-photon hyperentanglement in polarization-time-bin DOFs. Without loss of generality, we assume that $m$ is even. Then the quantum state of successfully generated $m$-photon hyperentanglement can be represented as Eq. (\ref{even}). Similarly, we will provide an in-depth analysis of multi-photon events in the context of multi-photon hyperentanglement in polarization-time-bin DOFs in the appendix.

\begin{figure}[h]
  \centering
 \includegraphics[scale=0.5,angle=0]{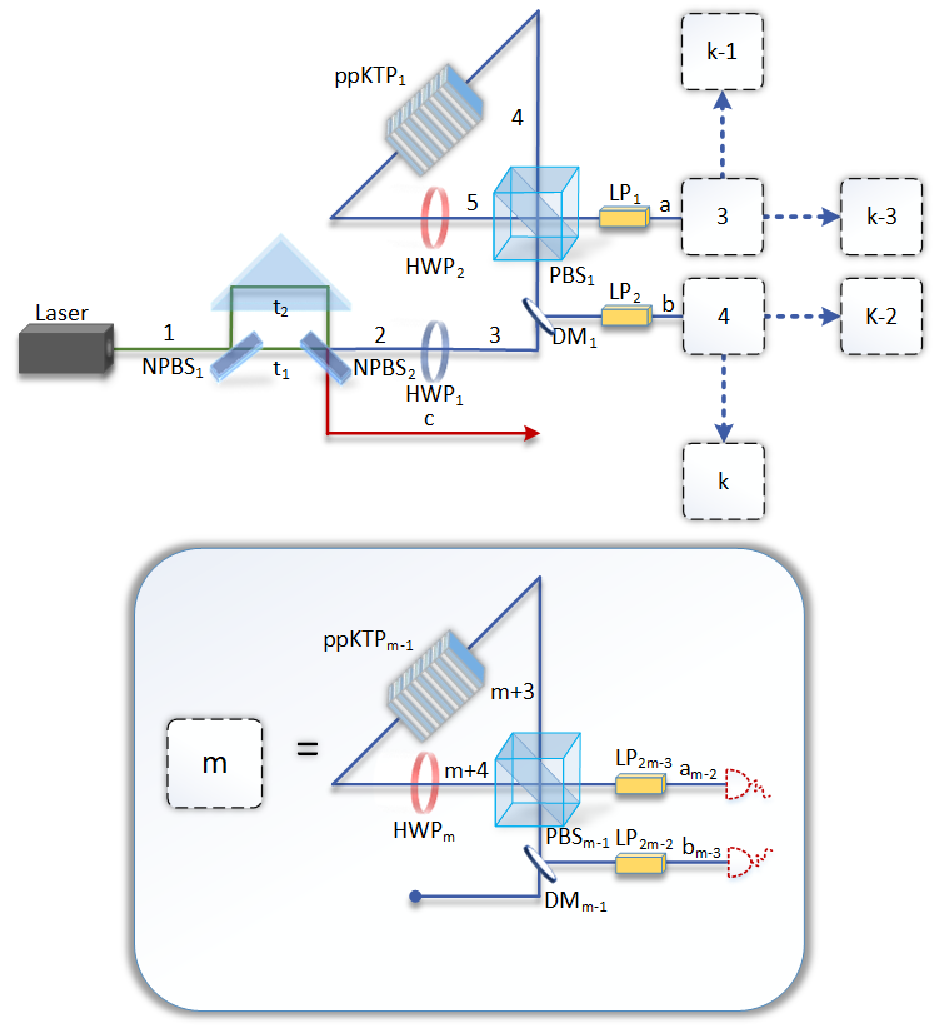}
  \caption{The schematic illustrates the generation of $m$-photon hyperentanglement in polarization-time-bin DOFs. HWP$_1$: $22.5^{\circ}$ half wave plate; HWP$_{2}$: $45^{\circ}$ half wave plate; NPBS: non-polarizing beam splitter; DM: dichroic mirror; PBS: polarizing beam splitter; ppKTP: periodically poled potassium titanyI phosphate; LP: long pass filter. The red dashed detector indicates that only the final cascade generates photons that may require detection.}
\end{figure}

\begin{eqnarray}\label{even}
  &&\frac{1}{(\sqrt{2})^{2m-4}}[(|HV\ldots VH\rangle+|VH\ldots HV\rangle) \nonumber\\
  &&\otimes(|t_1t_1\ldots t_1t_1\rangle+|t_2t_2\ldots t_2t_2\rangle)]_{b_1a_1 \ldots b_{m-2}a_{m-2}}.\nonumber\\
\end{eqnarray}

\begin{figure}[!htp]
  \centering
  \includegraphics[scale=0.9,angle=0]{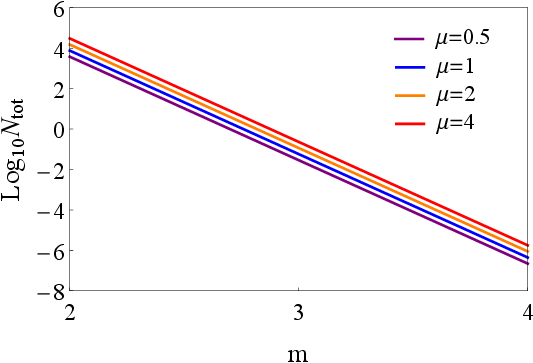}
  \caption{The function relationship between the number of generated hyperentangled pairs, mean photon number, and m. Here, we assume that the repetition rate of the pulses is $10^9$ Hz and the ppKTP efficiency is $7.6*10^{-6}$. }
\end{figure}

\section{Discussion and conclusion}
We have provided a detailed description of the protocol for generating polarization-space mode and polarization-time-bin cascaded hyperentangled states. Under actual experimental conditions, the laser source may emits multi-photon events with a certain probability.  In actual experiments, the repetition frequency of the laser lies between $10^6$ and $10^9$ Hz \cite{Kwiat,Hz1,Hz2,Hz3}. We can calculate the number of generated hyperentangled pairs  (see the Appendix for more details), assuming that the repetition rate of pulses is $10^9$ Hz and the ppKTP efficiency is $7.6*10^{-6}$ \cite{cascade1,cascade2}, as
\begin{eqnarray}\label{Ntot}
  N_{tot} &=& F*\sum_{n=0}^{\infty}p(n)*P_{n}^{m}(tot)_{succ} \nonumber\\
  &=&F*(1-e^{-\mu p_s^{(m-1)}}).
\end{eqnarray}
Here $F$ denotes the repetition frequency of laser. $p(n) = \frac{e^{-\mu}\mu^{n}}{n!}$ is the probability for the $n$-photon emission, which follows the Poisson distribution with the mean photon number $\mu$ \cite{poisson}. $N_{tot}$ is the amount of the generated hyperentanglement per second. To obtain a more intuitive conclusion, we took the logarithm of variable $N_{tot}$. As shown in Fig. 7, the horizontal axis represents generating $m$-photon hyperentanglement, and the vertical axis represents the number of hyperentangled pairs generated. When repetition frequency of laser and ppKTP are given, the number of hyperentangled pairs generated increases with the increase in the average photon number of the laser source. This is because multi-photon events provide positive contribution. Specifically, if the repetition rate of the laser is $10^9$ Hz and the ppKTP efficiency is $7.6*10^{-6}$, we can respectively get three-photon hyperentanglement approximately $2.89*10^{-2}$ pairs per second for $\mu =0.5$, $5.78*10^{-2}$ pairs per second for $\mu = 1$, $1.16*10^{-1}$ pairs per second for $\mu = 2$, and $2.31*10^{-1}$ pairs per second for $\mu = 4$. Due to the positive gain from multi-photon events, the number of generated hyperentangled pairs increases with a higher mean photon number. However, in practical experiments, a larger average photon number may lead to crystal breakdown and device damage. Therefore, the specific parameter requirements need to be determined based on practical considerations. From Fig. 7, we can also observe that as the value of $m$ increases, the generated photon pairs decrease. This is evident because a larger $m$ value implies a lower probability of successful cascading. If the repetition rate of the laser is $10^9$ Hz, the efficiency of ppKTP is $7.6*10^{-6}$ and $\mu =1$, the number of the generation of three-photon and four-photon hyperentanglement after cascading can reach approximately $5.78*10^{-2}$ and $4.44*10^{-7}$ pairs per second, respectively.
For our cascaded generation of $m$-photon hyperentanglement source, we can represent the state as follows,
\begin{eqnarray}
  |Cas\rangle_{m} &=& Pr(m,0)|0\rangle + Pr(m,1)|\phi\rangle + Pr(m,2)|\phi\rangle^{\otimes2} \nonumber\\
  &&+\cdots + Pr(m,n)|\phi\rangle^{\otimes n}.
\end{eqnarray}
Here, $|0\rangle$, $|\phi\rangle$, $|\phi\rangle^{\otimes2}$ and $|\phi\rangle^{\otimes n}$ represent the generation of zero pair, one pair, two pairs, and $n$ pairs of $m$-photon hyperentanglement, respectively. $|\phi\rangle$ is the $m$-photon hyperentanglement like Eqs. (\ref{desired}) and (\ref{even}).

Despite the preparation of entanglement has achieved significant advancements \cite{review1, review2, review3, review4}, the preparation of multi-photon hyperentangled states remains some challenges in current experimental conditions, demanding highly precise control to ensure entanglement between photons. For instance, the SPDC process necessitates satisfying phase matching conditions, and the requirements in the cascaded SPDC source scenario may be even more stringent. Taking Fig. 4 as an example, in the nonlinear crystal ppKTP, a pump photon with a frequency of $\omega_p$ will, with a relatively low probability, undergo down-conversion, splitting into a pair of twin photon with frequencies $\omega_1$ and $\omega_2$ respectively. Clearly, this process must satisfy energy conservation: $\hbar\omega_p = \hbar\omega_1 + \hbar\omega_2$. Here, $\hbar$ is the reduced Planck constant. Subsequently, the photon with frequency $\omega_1$ will, again with a certain probability, undergo SPDC, splitting into another pair of twin photon with $\omega_3$ and $\omega_4$. This cascaded SPDC process naturally satisfy energy conservation: $\hbar\omega_p = \hbar\omega_2 + \hbar\omega_3 + \hbar\omega_4$. From the Ref. \cite{cascade3}, we can obtain a simple expression for the frequency-space of this cascaded SPDC source.
\begin{eqnarray}
&&\Phi_{C_3} \approx \int_{\omega_2}\int_{\omega_3}G_1(\omega_2, \omega_p - \omega_2) G_2(\omega_3, \omega_p - \omega_2 - \omega_3) \nonumber\\
&&a_1^{\dag}(\omega_2)a_2^{\dag}(\omega_3)a_3^{\dag}(\omega_p - \omega_2 - \omega_3)|0\rangle d\omega_2d\omega_3.
\end{eqnarray}
Here, $G_1(\omega_2, \omega_p - \omega_2)$ and $G_2(\omega_3, \omega_p - \omega_2 - \omega_3)$ represent the joint spectral functions generated by phase matching conditions in the first and second ppKTP crystals respectively. $a_1^{\dag}(\omega_2)$, $a_2^{\dag}(\omega_3)$ and $a_3^{\dag}(\omega_p - \omega_2 - \omega_3)$ are the creation operators. Obviously, this cascaded approach can be extended to more photons (although the probability of successful implementation will be very small). As shown in Fig. 6, the entire cascaded SPDC process also needs to satisfy the energy conservation $\hbar\omega_p = \hbar\omega_{2^i-1} + \hbar\omega_{2^i}+\cdots+\hbar\omega_{2^{i+1}-2}$. Note that for the sake of convenience, we provide the energy conservation condition using the example of generating $2^i$ entangled photons. Here, $\omega_{2^i-1}, \omega_{2^i}, \cdots, \omega_{2^{i+1}-2}$ represent the frequencies of the split photons after SPDC. As mentioned in \cite{cascade3}, we can also provide a simplified expression for an $n$-photon state in frequency-space.
\begin{eqnarray}
&&\Phi_{C_{2^{i}}} \approx \int_{\omega_{2^i-1}}\cdots\int_{\omega_{2^{i+1}-3}}G_1(\omega_1, \omega_2) G_2(\omega_3, \omega_4) \nonumber\\
&&\times G_i(\omega_{2^{i+1}-3}, \omega_{2^{i+1}-2}) a_1^{\dag}(\omega_{2^i-1})a_2^{\dag}(\omega_{2^i})\cdots  \nonumber\\
&&a_{2^i}^{\dag}(\omega_{2^{i+1}-2})|0\rangle d\omega_{2^i-1}\cdots d\omega_{2^{i+1}-2},
\end{eqnarray}
where $G_i(\omega_{2^{i+1}-3}$, $\omega_{2^{i+1}-2})$ represents the joint spectral functions generated by phase matching conditions in the $i$-th ppKTP crystal. $a_{2^i}^{\dag}(\omega_{2^{i+1}-2})$ means the creation operator for the $i$-th photon.

During the preparation of entanglement, the birefringence effect of the crystal will introduce a relative phase between the down-converted photons. Taking Eq. (\ref{3}) as an example, $|HV\rangle$ and $|VH\rangle$ will have a relative phase $\theta$, directly resulting in the prepared quantum state not being an ideally maximally entangled state. Fortunately, as early as 1995, Zeilinger \emph{et al.} showed that by using of an additional birefringent phase shifter or by slightly rotating the converting crystal itself, the value of $\theta$ can be adjusted as desired, e.g., set to 0 or $\pi$ \cite{phase-difference1}. Ursin \emph{et al.} also pointed out that it is possible to compensate for the phase difference caused by the different group velocities of the pump light and down-converted photons in ppKTP by using a dual-wavelength HWP \cite{PF2}. Therefore, birefringent devices can be employed to achieve entangled states with a relative phase of 0 as required in Eq. (\ref{3}), and a similar approach can be applied to realize other cascade quantum states without elaborating further here.

The efficiency of conversion under non-linear crystals is also a major factor affecting the generation of entanglement. In the three-photon entangled state generation using cascaded photons directly from the source as proposed by H\"{u}bel \emph{et al.}, it was noted that the down-conversion efficiency of SPDC is extremely low \cite{cascade1}. In nonlinear crystal barium borate (BBO), the down-conversion efficiency for each pump photon can only reach $10^{-11}$ \cite{-11}. With the development of nonlinear optics, materials such as periodically poled lithium niobate (PPLN) and ppKTP have improved the efficiency to $10^{-9}$ \cite{-9}. By introducing waveguides, the down-conversion efficiency can be further enhanced to $10^{-6}$ \cite{-6}. Moreover, in the cascaded SPDC source proposed by Hamel \emph{et al.} for generating three-photon polarization entanglement, the authors indicated that the down-conversion efficiency can reach $(6.9\pm0.7)*10^{-6}$ \cite{cascade2} and $(1\pm0.1)*10^{-6}$ \cite{cascade2add}, respectively. In addition, the preparation of three-photon time-energy entanglement has also been proposed and experimentally demonstrated by researchers \cite{cascade3}. Although the above approaches are associated with low counting rate, it is believed that the efficiency of down-conversion will continue to improve with advancements in research technology and an increase in researchers' expertise, thereby promoting the increase of entanglement quantity.

Brightness characterizes the rate of entanglement generation. Numerous factors influence brightness, including the efficiency of down-conversion in the crystal, transmission losses, device losses, detector efficiency, and dark counts. Taking Fig. 1 as an example, assuming the coincident counts for spatial modes $c_1$, $c_2$, and $b_2$ are denoted as $C_{c_1c_2b_2}$, and for spatial modes $d_1$, $d_2$, and $a_2$ are denoted as $C_{d_1d_2a_2}$. Then, the brightness of the source can be expressed as $C_{c_1c_2b_2}+C_{d_1d_2a_2}$. Consequently, in the scenario depicted in Fig. 3, the brightness of the source can be analogously represented as $C_{d_{2m-4}d_{2m-3}\ldots d_2d_1}+C_{c_{2m-4}c_{2m-3}\ldots c_2c_1}$. Here, $C_{d_{2m-4}d_{2m-3}\ldots d_2d_1}$ and $C_{c_{2m-4}c_{2m-3}\ldots c_2c_1}$ represent the coincident counts for spatial modes $d_{2m-4}, d_{2m-3}\ldots d_2, d_1$ and $c_{2m-4}, c_{2m-3}\ldots c_2, c_1$, respectively.

In conclusion, we proposed the protocols of direct generation of three- and four-photon hyperentanglement with cascaded downcoversion, and the hyperentangled states are encoded in polarization-spatial modes and polarization-time-bin degrees of freedom, respectively. We also extended such approach to multi-photon hyperentangled states. The most advantage of these protocols are they do not relied on the post-selection strategy, and the produced states are the desired multi-photon hyperentangled state. This work has the potential to demonstrate
that combining multiparticle entanglement with multiple DOFs can provide an efficient route to increase both the number of effective qubits and capacity of future quantum communication and quantum network.

\section*{Appendix}
\setcounter{equation}{0}
\setcounter{subsection}{0}
\renewcommand{\theequation}{A\arabic{equation}}
This appendix provides additional computations concerning the generation probability of multi-photon hyperentanglement, aiming to support and extend the quantitative analyses presented in the main text. The following detailed descriptions of additional calculations contribute to a more comprehensive understanding of the research outcomes presented in this paper.

We start by illustrating the successful generation of two-photon hyperentanglement in polarization-spatial-mode DOFs in the case of a two-photon event. In cases leading to the creation of a pair of three-photon hyperentanglement, two distinct scenarios arise. In the first scenario, one photon is selected from the two-photon event and successfully undergoes splitting on ppKTP$_1$ and ppKTP$_2$, while the other photon fails to split on the ppKTP$_1$. In the second scenario, one photon is chosen from the two-photon event and successfully undergoes splitting on ppKTP$_1$ and ppKTP$_2$, but the other photon successfully splits on ppKTP$_1$ and fails to split on ppKTP$_2$. This probability can be expressed by the following equation.
\begin{eqnarray}\label{p231}
  P_{2}^{3}(1)_{succ} &=& C_{2}^{1}p_s^{2}(1-p_s)+C_{2}^{1}p_s^{2}p_s(1-p_s),
\end{eqnarray}
where $P_{2}^{3}(1)_{succ}$ represents the probability of a two-photon event successfully generating one pair of three-photon hyperentanglement. $C$ denotes the combination calculation. $p_s$ is the probability of successful splitting on ppKTP. Subsequent success probabilities have similar meanings, and they are not further elaborated in this paper.

In the case of generating two pairs of three-photon hyperentanglement, both photons must undergo successful splitting. Therefore, it can be expressed as follows.
\begin{eqnarray}\label{p232}
  P_{2}^{3}(2)_{succ} &=& p_s^{2}p_s^{2}.
\end{eqnarray}

In this way, the total probability of successfully generating three-photon hyperentanglement in two-photon event can be expressed as follows.
\begin{eqnarray}\label{2-3-tot}
  P_{2}^{3}(tot)_{succ} &=& P_{2}^{3}(1)_{succ}+P_{2}^{3}(2)_{succ} \nonumber\\
  &=&C_{2}^{1}p_s^{2}(1-p_s)+C_{2}^{1}p_s^{2}p_s(1-p_s)+p_s^{2}p_s^{2} \nonumber\\
  &=&2p_s^{2}-p_s^{4}.
\end{eqnarray}

We can verify the accuracy of our calculations by backward computing the probabilities of failure. Specifically, we categorize cases of failure in a two-photon event (where three-photon hyperentanglement is not generated) into three cases. Case 1 involves both photons not undergoing splitting as they pass through the ppKTP$_1$. It can be expressed as follows.

\begin{eqnarray}
  P_{2}^{3}(0)_{fail} &=& (1-p_s)(1-p_s).
\end{eqnarray}
Here, $P_{2}^{3}(0)_{fail}$ represents the probability of failure generating three-photon hyperentanglement in two-photon event, where the failure occurs due to the non-simultaneous splitting of the two photons on ppKTP. The number inside the parentheses indicates the events of failure in multi-photon cases, representing the scenario where a certain number of photons simultaneously underwent splitting on ppKTP$_1$ but failed to generate three-photon hyperentanglement successfully. The subsequent failure probabilities have similar meanings and will not be reiterated here.

Case 2 is when one of the photons successfully undergoes splitting on ppKTP$_1$, but this photon fails to split on ppKTP$_2$. This can be expressed as follows.
\begin{eqnarray}
  P_{2}^{3}(1)_{fail} &=&C_{2}^{1}p_s(1-p_s)(1-p_s).
\end{eqnarray}

Case 3 entails both photons successfully undergoing splitting on ppKTP$_1$ but failing to split on ppKTP$_2$. The probability can be represented as

\begin{eqnarray}
  P_{2}^{3}(2)_{fail} &=& p_s^{2}(1-p_s)^{2}.
\end{eqnarray}

In this way, the total failure probability in a two-photon event is given by the following expression.
\begin{eqnarray}
  P_{2}^{3}(tot)_{fail} &=& P_{2}^{3}(0)_{fail}+P_{2}^{3}(1)_{fail}+P_{2}^{3}(2)_{fail}  \nonumber\\
  &=&(1-p_s)(1-p_s)+C_{2}^{1}p_s(1-p_s)(1-p_s) \nonumber\\
  &&+p_s^{2}(1-p_s)^{2} \nonumber\\
  &=&1-2p_s^{2}+p_s^{4}.
\end{eqnarray}
It can be observed that the sum of the success probability and the failure probability is equal to 1, validating the accuracy of our calculations.

Similarly, we conduct numerical calculations for the case of generating three-photon hyperentanglement in a three-photon event. The three-photon event can result in the creation of one pair, two pairs, and three pairs of three-photon hyperentanglement. In the case of generating one pair of three-photon hyperentanglement, there are three scenarios. Scenario 1 involves selecting one photon from the three photons, and this photon successfully undergoes splitting on ppKTP$_1$ and  ppKTP$_2$, while the remaining two photons fail to split on ppKTP$_1$. Scenario 2 includes selecting one photon from the three photons, and this photon successfully undergoes splitting on ppKTP$_1$ and the ppKTP$_2$, while the remaining two photons undergo successful splitting on ppKTP$_1$ but fail to split on ppKTP$_2$. Scenario 3 comprises selecting one photon from the three photons, and this photon successfully undergoes splitting on ppKTP$_1$ and  ppKTP$_2$, while one of the remaining two photons undergoes successful splitting on ppKTP$_1$ but fails to split on ppKTP$_2$, and the last remaining photon fails to split on ppKTP$_1$. Therefore, we can calculate the probabilities for these three scenarios as follows.

\begin{eqnarray}\label{p331}
  P_{3}^{3}(1)_{succ} &=& C_{3}^{1}p_s^{2}(1-p_s)^{2}+C_{3}^{1}p_s^{4}(1-p_s)^{2} \nonumber\\
  &&+C_{3}^{1}p_s^{2}C_{2}^{1}p_s(1-p_s)^{2}.
\end{eqnarray}

There are two scenarios in the case of generating two pairs of three-photon hyperentanglement. Scenario 1 involves selecting two photons from the three photons, and these two photons successfully undergo splitting on ppKTP$_1$ and ppKTP$_2$, while the remaining photon fails to split on ppKTP$_1$. Scenario 2 includes selecting two photons from the three photons, and these two photons successfully undergo splitting on ppKTP$_1$ and ppKTP$_2$, while the remaining photon successfully undergoes splitting on ppKTP$_1$ but fails to split on ppKTP$_2$. The probability of generating two pairs of three-photon hyperentanglement in this case is given by the following expression.
\begin{eqnarray}\label{p332}
  P_{3}^{3}(2)_{succ} &=& C_{3}^{2}p_s^{4}(1-p_s)+C_{3}^{2}p_s^{5}(1-p_s).
\end{eqnarray}

In the scenario of generating three pairs of three-photon hyperentanglement, there is only one scenario where all three photons successfully undergo splitting on ppKTP$_1$ and ppKTP$_2$. The probability in this scenario can be expressed as
\begin{eqnarray}
  P_{3}^{3}(3)_{succ} &=& p_s^{6}.
\end{eqnarray}

In this way, the probability of successfully generating three-photon hyperentanglement in a three-photon event can be expressed as follows.
\begin{eqnarray}\label{3-3-tot}
  P_{3}^{3}(tot)_{succ} &=& P_{3}^{3}(1)_{succ}+P_{3}^{3}(2)_{succ}+P_{3}^{3}(3)_{succ}  \nonumber\\
  &=&3p_s^{2}-3p_s^{4}+p_s^{6}.
\end{eqnarray}

Similarly, we can validate our calculations by backward computing the probabilities of failure. In the case of failure, there are four scenarios. Scenario 1 is when all three photons fail to split on ppKTP$_1$. This can be expressed as follows.
\begin{eqnarray}
  P_{3}^{3}(0)_{fail} &=& (1-p_s)^{3}.
\end{eqnarray}

Scenario 2 involves one photon successfully undergoing splitting on ppKTP$_1$, but this photon fails to split on ppKTP$_2$. The probability in this case is given by the following expression.
\begin{eqnarray}
  P_{3}^{3}(1)_{fail} &=& C_{3}^{1}p_s(1-p_s)^{3}.
\end{eqnarray}

Scenario 3 occurs when two photons successfully split on ppKTP$_1$, but both of these photons fail to split on ppKTP$_2$. The probability in this scenario is expressed as follows.
\begin{eqnarray}
  P_{3}^{3}(2)_{fail} &=& C_{3}^{2}p_s^{2}(1-p_s)^{3}.
\end{eqnarray}

Scenario 4 entails successful splitting of all three photons on ppKTP$_1$, while none of these photons successfully splits on ppKTP$_2$. The probability in this case is given by the following expression.
\begin{eqnarray}
  P_{3}^{3}(3)_{fail} &=& p_s^{3}(1-p_s)^{3}.
\end{eqnarray}

Naturally, the total failure probability is given by the following expression.
\begin{eqnarray}
  P_{3}^{3}(tot)_{fail} &=&  P_{3}^{3}(0)_{fail}+P_{3}^{3}(1)_{fail}+P_{3}^{3}(2)_{fail} \nonumber\\
  &&+P_{3}^{3}(3)_{fail}  \nonumber\\
  &=& 1-3p_s^{2}+3p_s^{4}-p_s^{6}.
\end{eqnarray}

As expected, the sum of the failure probability and the success probability is equal to 1, validating the accuracy of the calculations for the generation of three-photon hyperentanglement in a three-photon event.

In the preceding sections, we calculated the probabilities of generating three-photon hyperentanglement in dual-photon and tri-photon events. We validated the results by computing the probabilities of failure. In practical laser sources, events involving more than three photons may occur. Now, we present the general formula and computation process for the n-photon event.

Here, we calculate the probability of success by computing the probability of failure, as it is relatively straightforward. In the case of an $n$-photon event successfully generating three-photon hyperentanglement, there are $n+1$ scenarios. Scenario 1 is when all $n$ photons fail to split on ppKTP$_1$, expressed as follows.
\begin{eqnarray}
  P_{n}^{3}(0)_{fail} &=& (1-p_s)^{n}.
\end{eqnarray}

Scenario 2 involves selecting one photon from the $n$ photons, which successfully undergoes splitting on ppKTP$_1$ while this photon fails to split on ppKTP$_2$. It can be expressed as follows.
\begin{eqnarray}
  P_{n}^{3}(1)_{fail} &=& C_{n}^{1}p_s(1-p_s)^{n-1}(1-p_s).
\end{eqnarray}

Scenario 3 refers to selecting two photons from the $n$ photons, where both successfully undergo splitting on ppKTP$_1$ while these two photons fail to split on ppKTP$_2$. The probability in this case can be expressed as follows.
\begin{eqnarray}
  P_{n}^{3}(2)_{fail} &=& C_{n}^{2}p_s^{2}(1-p_s)^{n-2}(1-p_s)^{2}.
\end{eqnarray}

Continuing in this manner, we can write the probability for scenario $i+1$. This represents selecting $i$ photons from the $n$ photons, where these $i$ photons successfully undergo splitting on ppKTP$_1$ and fail to split on ppKTP$_2$. Here, $i\in \{0,1,2\ldots n \}$. The general formula can be expressed as follows.
\begin{eqnarray}
  P_{n}^{3}(i)_{fail} &=& C_{n}^{i}p_s^{i}(1-p_s)^{n-i}(1-p_s)^{i}.
\end{eqnarray}

By summing up the general formula for the probability of failure, we can obtain the overall probability of success, given by the following expression.
\begin{eqnarray}\label{sum1}
  P_{n}^{3}(tot)_{succ} &=& 1-\sum_{i=0}^{n}P_{n}^{3}(i)_{fail} \nonumber\\
  &=&1-\sum_{i=0}^{n}C_{n}^{i}p_s^{i}(1-p_s)^{n-i}(1-p_s)^{i}  \nonumber\\
  &=&1-(1-p_s)^{n}(1+p_s)^{n}=1-(1-p_s^{2})^{n}. \nonumber\\
\end{eqnarray}
By substituting $n=2$ and $n=3$ into Eq. (\ref{sum1}), we can verify the correctness of Eqs. (\ref{2-3-tot}) and (\ref{3-3-tot}).

So far, we have provided the probability of successfully generating three-photon hyperentanglement in the case of multi-photon events. For clarity, we will proceed to calculate the probability of successfully generating four-photon hyperentanglement in multi-photon events. Through these two examples, we aim to derive the probability of successfully generating $m$-photon hyperentanglement in multi-photon events. Clearly, we should start by calculating the probability of successfully generating four-photon hyperentanglement in a two-photon event, which includes two cases: the creation of one pair of four-photon hyperentanglement and two pairs of four-photon hyperentanglement.

The first case consists of two scenarios. In Scenario 1, one photon from the two-photon event successfully undergoes splitting on ppKTP$_1$, ppKTP$_2$ and ppKTP$_3$, while the remaining photon fails to split on ppKTP$_1$. In Scenario 2, one photon from the two-photon event successfully undergoes splitting on ppKTP$_1$, ppKTP$_2$ and ppKTP$_3$, while the remaining photon successfully undergoes splitting on ppKTP$_1$ but fails to split on ppKTP$_2$ and ppKTP$_3$ simultaneously. In this case, the probability of the first case is given by the following expression.
\begin{eqnarray}\label{p241}
  P_{2}^{4}(1)_{succ} &=& C_{2}^{1}p_s^{3}(1-p_s)+C_{2}^{1}p_s^{4}(1-p_s^{2}).
\end{eqnarray}

The second case is obvious that both photons in the two-photon event must successfully undergo splitting on ppKTP$_1$, ppKTP$_2$ and ppKTP$_3$, which is expressed as follows.
\begin{eqnarray}\label{p242}
  P_{2}^{4}(2)_{succ} &=& p_s^{6}.
\end{eqnarray}

In this way, the probability of successfully generating four-photon hyperentanglement in a two-photon event can be expressed as follows.
\begin{eqnarray}\label{2-4-tot}
  P_{2}^{4}(tot)_{succ} &=& P_{2}^{4}(1)_{succ}+P_{2}^{4}(2)_{succ}=2p_s^{3}-p_s^{6}. \nonumber\\
\end{eqnarray}

Similar to the consideration in the previous discussion, we can verify the correctness of the probability of success by calculating the probability of failure. This involves three scenarios. In Scenario 1, both photons in the two-photon event fail to split on ppKTP$_1$. In Scenario 2, one photon from the two-photon event successfully undergoes splitting on ppKTP$_1$, but this photon fails to split on ppKTP$_2$ and ppKTP$_3$ simultaneously. In Scenario 3, both photons from the two-photon event successfully undergo splitting on ppKTP$_1$, but these two photons fail to split on ppKTP$_2$ and ppKTP$_3$ simultaneously. The probabilities for these three scenarios can be represented by Eqs. (\ref{fail2-4-0}), (\ref{fail2-4-1}) and (\ref{fail2-4-2}), respectively.
\begin{eqnarray}\label{fail2-4-0}
  P_{2}^{4}(0)_{fail} &=& (1-p_s)^{2}.
\end{eqnarray}

\begin{eqnarray}\label{fail2-4-1}
  P_{2}^{4}(1)_{fail} &=& C_{2}^{1}p_s(1-p_s)(1-p_s^{2}).
\end{eqnarray}

\begin{eqnarray}\label{fail2-4-2}
  P_{2}^{4}(2)_{fail} &=& p_s^{2}(1-p_s^{2})^{2}.
\end{eqnarray}

In this case, the probability of not successfully generating four-photon hyperentanglement in a two-photon event is given by the following expression.
\begin{eqnarray}
  P_{2}^{4}(tot)_{fail} &=& P_{2}^{4}(0)_{fail}+P_{2}^{4}(1)_{fail}+P_{2}^{4}(2)_{fail}  \nonumber\\
  &=&(1-p_s)^{2}+C_{2}^{1}p_s(1-p_s)(1-p_s^{2}) \nonumber\\
  &&+p_s^{2}(1-p_s^{2})^{2}   \nonumber\\
  &=&1+p_s^{6}-2p_s^{3}.
\end{eqnarray}

We can observe that this is consistent with the fact that the sum of the probability of failure and the probability of success equals 1. The successful generation of four-photon hyperentanglement in a three-photon event includes three cases: the creation of one pair, two pairs, and three pairs of four-photon hyperentanglement.

In the case of creating one pair, there are three scenarios. In Scenario 1, one photon from the three-photon event successfully undergoes splitting on ppKTP$_1$, ppKTP$_2$, and ppKTP$_3$, while the remaining photon fail to split on ppKTP$_1$. In Scenario 2, one photon from the three-photon event successfully undergoes splitting on ppKTP$_1$, ppKTP$_2$ and ppKTP$_3$, while one of the remaining photons successfully undergoes splitting on ppKTP$_1$ but fails to split on ppKTP$_2$ and ppKTP$_3$ simultaneously. In Scenario 3, one photon from the three-photon event successfully undergoes splitting on ppKTP$_1$, ppKTP$_2$, and ppKTP$_3$, while the remaining two photons successfully undergo splitting on ppKTP$_1$, but these two photons fail to split on ppKTP$_2$ and ppKTP$_3$ simultaneously. In this case, the probability of creating one pair of four-photon hyperentanglement is given by the following expression.
\begin{eqnarray}\label{p341}
  P_{3}^{4}(1)_{succ} &=& C_{3}^{1}p_s^{3}(1-p_s)^{2}+C_{3}^{1}p_s^{3}C_{2}^{1}p_s(1-p_s^{2})(1-p_s) \nonumber\\
  &&+ C_{3}^{1}p_s^{5}(1-p_s^{2})^{2}.
\end{eqnarray}

In the case of creating two pairs of four-photon hyperentanglement, there are two scenarios. Scenario 1 involves three photons, where two photons successfully undergo splitting on ppKTP$_1$, ppKTP$_2$, and ppKTP$_3$, while the remaining photon fails to split on ppKTP$_1$. Scenario 2 involves three photons, where two photons successfully undergo splitting on ppKTP$_1$, ppKTP$_2$, and ppKTP$_3$, while the remaining photon successfully undergoes splitting on ppKTP$_1$ but fails to split on ppKTP$_2$ and ppKTP$_3$ simultaneously. In this case, the probability of creating two pairs of four-photon hyperentanglement is given by the following expression.

\begin{eqnarray}\label{p342}
  P_{3}^{4}(2)_{succ} &=& C_{3}^{2}p_s^{6}(1-p_s)+C_{3}^{2}p_s^{7}(1-p_s^{2}).
\end{eqnarray}

The case of creating three pairs of four-photon hyperentanglement is straightforward, as it requires all three photons to successfully undergo splitting on ppKTP$_1$, ppKTP$_2$, and ppKTP$_3$. In this case, the probability of creating three pairs of four-photon hyperentanglement can be expressed as follows.

\begin{eqnarray}
  P_{3}^{4}(3)_{succ} &=& p_s^{9}.
\end{eqnarray}

The total probability of a three-photon event successfully generating four-photon hyperentanglement can be written as
\begin{eqnarray}\label{3-4-tot}
  P_{3}^{4}(tot)_{succ} &=& P_{3}^{4}(1)_{succ}+P_{3}^{4}(2)_{succ}+P_{3}^{4}(3)_{succ} \nonumber\\
  &=&C_{3}^{1}p_s^{3}(1-p_s)^{2}+C_{3}^{1}p_s^{3}C_{2}^{1}p_s(1-p_s^{2}) \nonumber\\
  &&\times(1-p_s)+C_{3}^{1}p_s^{5}(1-p_s^{2})^{2} \nonumber\\
  &+&C_{3}^{2}p_s^{6}(1-p_s)+C_{3}^{2}p_s^{7}(1-p_s^{2})+ p_s^{9} \nonumber\\
  &=&3p_s^{3}-3p_s^{6}+p_s^{9}.
\end{eqnarray}

Similarly, we calculate the probability of failure separately to confirm the correctness of the probability of success. This involves four scenarios. In Scenario 1, all three photons fail to split on ppKTP$_1$. In Scenario 2, one photon from the three-photon event successfully undergoes splitting on ppKTP$_1$, but this photon fails to split on ppKTP$_2$ and ppKTP$_3$ simultaneously. In Scenario 3, two photons from the three-photon event successfully undergo splitting on ppKTP$_1$, but these two photons fail to split on ppKTP$_2$ and ppKTP$_3$ simultaneously. In Scenario 4, all three photons from the three-photon event successfully undergo splitting on ppKTP$_1$, but these photons fail to split on ppKTP$_2$ and ppKTP$_3$ simultaneously. The probabilities for these four scenarios can be represented by Eqs. (\ref{fail3-4-0}), (\ref{fail3-4-1}), (\ref{fail3-4-2}), and (\ref{fail3-4-3}), respectively.
\begin{eqnarray}\label{fail3-4-0}
  P_{3}^{4}(0)_{fail} &=& (1-p_s)^{3}.
\end{eqnarray}

\begin{eqnarray}\label{fail3-4-1}
  P_{3}^{4}(1)_{fail} &=& C_{3}^{1}p_s(1-p_s^{2})(1-p_s)^{2}.
\end{eqnarray}

\begin{eqnarray}\label{fail3-4-2}
  P_{3}^{4}(2)_{fail} &=& C_{3}^{2}p_s^{2}(1-p_s^{2})^{2}(1-p_s).
\end{eqnarray}

\begin{eqnarray}\label{fail3-4-3}
  P_{3}^{4}(3)_{fail} &=& p_s^{3}(1-p_s^{2})^{3}.
\end{eqnarray}

The failure probability of a three-photon event generating four-photon hyperentanglement is obtained from Eq. (\ref{fail3-4-tot}), confirming the correctness of our conclusion.
\begin{eqnarray}\label{fail3-4-tot}
  P_{3}^{4}(tot)_{fail} &=& P_{3}^{4}(0)_{fail}+ P_{3}^{4}(1)_{fail}+ P_{3}^{4}(2)_{fail}  \nonumber\\
  &&+P_{3}^{4}(3)_{fail}  \nonumber\\
  &=&1-3p_s^{3}+3p_s^{6}-p_s^{9}.
\end{eqnarray}

So far, we have provided the probabilities of generating four-photon hyperentanglement in two-photon and three-photon events. Afterwards, we will present the general formula for $n$-photon events and the total probability of generating four-photon hyperentanglement in multi-photon events. For ease of computation, we will illustrate using the probability of failure. This involves $n+1$ scenarios. In Scenario 1, none of the $n$ photons successfully undergo splitting on ppKTP$_1$. In Scenario 2, one photon from the $n$ photons successfully undergoes splitting on ppKTP$_1$, but this photon fails to split on ppKTP$_2$ and ppKTP$_3$ simultaneously. In Scenario 3, two photons from the n photons successfully undergo splitting on ppKTP$_1$, but these two photons fail to split on ppKTP$_2$ and ppKTP$_3$ simultaneously. The probabilities for these three scenarios can be represented by Eqs.  (\ref{failn-4-0}), (\ref{failn-4-1}), and (\ref{failn-4-2}), respectively.

\begin{eqnarray}\label{failn-4-0}
  P_{n}^{4}(0)_{fail} &=& (1-p_s)^{n}.
\end{eqnarray}

\begin{eqnarray}\label{failn-4-1}
  P_{n}^{4}(1)_{fail} &=& C_{n}^{1}p_s(1-p_s^{2})(1-p_s)^{n-1}.
\end{eqnarray}

\begin{eqnarray}\label{failn-4-2}
  P_{n}^{4}(2)_{fail} &=& C_{n}^{2}p_s^{2}(1-p_s^{2})^{2}(1-p_s)^{n-2}.
\end{eqnarray}

Continuing in this manner, we present the general formula for the probability in Scenario $i+1$, which can be represented as
\begin{eqnarray}
  P_{n}^{4}(i)_{fail} &=& C_{n}^{i}p_s^{i}(1-p_s^{2})^{i}(1-p_s)^{n-i}.
\end{eqnarray}

In this way, the probability of successfully generating four-photon hyperentanglement in an n-photon event can be expressed as follows.
\begin{eqnarray}\label{n-4-tot}
  P_{n}^{4}(tot)_{succ} &=& 1-\sum_{i=0}^{n}P_{n}^{4}(i)_{fail} \nonumber\\
  &=&1-\sum_{i=0}^{n}C_{n}^{i}p_s^{i}(1-p_s^{2})^{i}(1-p_s)^{n-i}  \nonumber\\
  &=& 1-(1-p_s^{3})^{n}.
\end{eqnarray}

It can be observed that Eq. (\ref{n-4-tot}) satisfies the conclusions of Eqs. (\ref{2-4-tot}) and (\ref{3-4-tot}), providing evidence of the correctness of our calculations.

So far, we have completed the calculations for the probability of successfully generating three- and four-photon hyperentanglement in $n$-photon events. Depending on the experimental requirements, researchers may also seek to generate hyperentanglement with more photons, such as five-photon hyperentanglement, six-photon hyperentanglement, or more. It is necessary to provide a general formula for the probability of successfully generating $m$-photon hyperentanglement in an $n$-photon event. Naturally, the calculations for successfully generating three- and four-photon hyperentanglements have already given us enough inspiration to address this issue. Specifically, for ease of understanding, we will also divide it into $n+1$ scenarios. In Scenario 1, none of the n photons successfully undergo splitting on ppKTP$_1$. In Scenario 2, one photon from the n photons successfully undergoes splitting on ppKTP$_1$, but this photon fails to split simultaneously on the remaining m-2 ppKTP crystals. In Scenario 3, two photons from the n photons successfully undergo splitting on ppKTP$_1$, but these two photons fail to split simultaneously on the remaining m-2 ppKTP crystals. The probabilities for the three scenarios mentioned above can be represented by Eqs. (\ref{failn-m-0}), (\ref{failn-m-1}), and (\ref{failn-m-2}), respectively.
\begin{eqnarray}\label{failn-m-0}
  P_{n}^{m}(0)_{fail} &=& (1-p_s)^{n}.
\end{eqnarray}

\begin{eqnarray}\label{failn-m-1}
  P_{n}^{m}(1)_{fail} &=& C_{n}^{1}p_s(1-p_s^{m-2})(1-p_s)^{n-1}.
\end{eqnarray}

\begin{eqnarray}\label{failn-m-2}
  P_{n}^{m}(2)_{fail} &=& C_{n}^{2}p_s^{2}(1-p_s^{m-2})^{2}(1-p_s)^{n-2}.
\end{eqnarray}

Continuing in this manner, the general formula for the probability of failure in Scenario $i+1$ can be expressed as follows.

\begin{eqnarray}
  P_{n}^{m}(i)_{fail} &=& C_{n}^{i}p_s^{i}(1-p_s^{m-2})^{i}(1-p_s)^{n-i}.
\end{eqnarray}

Thus, by summing up the probabilities of failure, we can ultimately obtain the total probability of successfully generating $m$-photon hyperentanglement in an $n$-photon event.
\begin{eqnarray}
  P_{n}^{m}(tot)_{succ} &=& 1-\sum_{i=0}^{n}P_{n}^{m}(i)_{fail} \nonumber\\
  &=&1-\sum_{i=0}^{n}C_{n}^{i}p_s^{i}(1-p_s^{m-2})^{i}(1-p_s)^{n-i}  \nonumber\\
  &=& 1-(1-p_s^{m-1})^{n}.
\end{eqnarray}

Thus, we can obtain Eq. (\ref{Ntot}) as follows,
\begin{eqnarray}
  N_{tot} &=& F*\sum_{n=0}^{\infty}p(n)*P_{n}^{m}(tot)_{succ} \nonumber\\
  &=&F*\sum_{n=0}^{\infty}\frac{e^{-\mu}\mu^{n}}{n!}*[1-(1-p_s^{m-1})^{n}] \nonumber\\
  &=&F*[\sum_{n=0}^{\infty}\frac{e^{-\mu}\mu^{n}}{n!}-\sum_{n=0}^{\infty}\frac{e^{-\mu}\mu^{n}}{n!}(1-p_s^{m-1})^n]  \nonumber\\
  &=&F*\{1-\sum_{n=0}^{\infty}\frac{e^{-\mu}[\mu(1-p_s^{m-1})]^{n}}{n!}\}  \nonumber\\
  &=&F*[1-e^{-\mu+\mu(1-p_s^{m-1})]} \nonumber\\
  &=&F*(1-e^{-\mu p_s^{(m-1)}}).
\end{eqnarray}

Combining Eqs. (\ref{p231}) and (\ref{p331}), we can generalize the probability of generating one pair of three-photon hyperentanglement in an $n$-photon event as follows,
\begin{eqnarray}\label{pn31}
  P_{n}^{3}(1)_{succ} &=& C_{n}^{1}p_s^{2}(1-p_s)^{n-1}+C_{n}^{1}p_s^{2}C_{n-1}^{1}p_s(1-p_s)^{n-1} \nonumber\\
  &&+C_{n}^{1}p_s^{2}C_{n-1}^{2}p_s^2(1-p_s)^{n-1}+\cdots \nonumber\\
  &&+C_{n}^{1}p_s^{2}C_{n-1}^{n-1}p_s^{n-1}(1-p_s)^{n-1} \nonumber\\
  &=&C_{n}^{1}p_s^{2}(1-p_s)^{n-1}(1+p_s)^{n-1} \nonumber\\
  &=&C_{n}^{1}p_s^{2}(1-p_s^2)^{n-1}.
\end{eqnarray}

Obtaining such results is expected, as Eq. (\ref{pn31}) represents an $n$-photon event where one photon successfully undergoes down-conversion on two crystals, resulting in one pair of hyperentangled photons, while the remaining n-1 photons fail to undergo down-conversion simultaneously on the two crystals.

Similarly, by combining Eqs. (\ref{p232}) and (\ref{p332}), we can derive the probability of generating two pairs of three-photon hyperentanglement in an $n$-photon event.
\begin{eqnarray}\label{pn32}
  P_{n}^{3}(2)_{succ} = C_{n}^{2}p_s^{4}(1-p_s^2)^{n-2}.
\end{eqnarray}

Afterwards, we can obtain the probabilities of generating one pair and two pairs of four-photon hyperentanglement in an $n$-photon event as well.
\begin{eqnarray}\label{pn41}
  P_{n}^{4}(1)_{succ} = C_{n}^{1}p_s^{3}(1-p_s^3)^{n-1}.
\end{eqnarray}

\begin{eqnarray}\label{pn42}
  P_{n}^{4}(2)_{succ} = C_{n}^{2}p_s^{6}(1-p_s^3)^{n-2}.
\end{eqnarray}

In this way, by combining Eqs. (\ref{pn31}), (\ref{pn32}), (\ref{pn41}) and (\ref{pn42}), we can ultimately obtain the probability of generating $r$ pairs of $m$-photon hyperentanglement in an $n$-photon event.
\begin{eqnarray}\label{pnmr}
  P_{n}^{m}(r)_{succ} = C_{n}^{r}p_s^{r(m-1)}(1-p_s^{m-1})^{n-r}.
\end{eqnarray}

If  considering a Poissonian distribution in laser source, the probability of generating $r$ pairs in the case of $m$-photon hyperentanglement can be represented as
\begin{eqnarray}
Pr(m,r) =&& \sum_{n=0}^{\infty}p(n)C_{n}^{r}p_s^{r(m-1)}(1-p_s^{m-1})^{n-r} \nonumber\\
=&&\sum_{n=0}^{\infty}e^{-\mu}\mu^{n}\frac{1}{r!(n-r)!}p_s^{r(m-1)}(1-p_s^{m-1})^{n-r} \nonumber\\
\xrightarrow{n-r=t}&&\sum_{t=0}^{\infty}e^{-\mu}\mu^{t}\mu^{r}\frac{1}{r!t!}p_s^{r(m-1)}[1-p_s^{(m-1)}]^{t} \nonumber\\
=&&\sum_{t=0}^{\infty}e^{-\mu}\frac{[\mu(1-p_s^{m-1})]^{t}}{t!}\frac{\mu^{r}}{r!}p_s^{r(m-1)} \nonumber\\
=&&e^{-\mu}\frac{\mu^{r}}{r!}e^{\mu(1-p_s^{m-1})}p_s^{r(m-1)} \nonumber\\
=&&\frac{\mu^{r}}{r!}e^{-\mu p_s^{m-1}}p_s^{r(m-1)}.
\end{eqnarray}

Take three-photon hyperentanglement as an example, we can obtain the ratio between two pairs and one pairs, which is $Pr(3,2)/Pr(3,1)=\frac{\mu p_s^{2}}{2}$. For $\mu=1$, $p_s=7.6*10^{-6}$, this ratio can reach about $2.88*10^{-11}$.

For our cascaded generation of $m$-photon hyperentanglement source, we can represent the state as follows,
\begin{eqnarray}
  |Cas\rangle_{m} &=& Pr(m,0)|0\rangle + Pr(m,1)|\phi\rangle + Pr(m,2)|\phi\rangle^{\otimes2} \nonumber\\
  &&+\cdots + Pr(m,n)|\phi\rangle^{\otimes n}.
\end{eqnarray}
Here, $|0\rangle$, $|\phi\rangle$, $|\phi\rangle^{\otimes2}$ and $|\phi\rangle^{\otimes n}$ represent the generation of zero pair, one pair, two pairs, and $n$ pairs of $m$-photon hyperentanglement, respectively. $|\phi\rangle$ is the $m$-photon hyperentanglement like Eqs. (\ref{desired}) and (\ref{even}).
\section*{Acknowledgement}
 We gratefully thank Cen-Xiao Huang, Chao Zhang, and Xiao-Ming Hu in University of Science and Technology of China for helpful discussion about the brightness, coincidence efficiency, and the crystal coherence of the generation protocols. This work is supported by the National Natural Science Foundation of China under Grant  Nos. 92365110, 12175106 and  11974189,  and Postgraduate Research \& Practice Innovation Program of Jiangsu Province under Grant No. KYCX23-1027.

\end{document}